%% file: arXiv_v2.tex
\documentclass[12pt]{article}
\pdfoutput=1 
\usepackage{graphicx}
\usepackage{subfig}
\usepackage{float}
\usepackage{amsmath}
\usepackage{amsmath,bm}
\usepackage{amssymb}
\usepackage{amsthm}
\usepackage{latexsym}
\usepackage{dcolumn}
\usepackage{geometry}
\usepackage{color}
\usepackage{mciteplus}
\usepackage{fancyhdr}
\usepackage{multicol}
\usepackage{multirow}
\usepackage{booktabs}
\usepackage{caption}
\usepackage{colortbl}
\usepackage{hyperref}
\usepackage{diagbox}
\usepackage{arydshln}
\usepackage[inline]{enumitem}
\usepackage{tabularx}

\newcolumntype{C}[1]{>{\centering\arraybackslash}p{#1}}

\definecolor{LightBlue}{rgb}{0.87, 0.94, 1}
\definecolor{LightRed}{rgb}{1, 0.85, 0.85}
\definecolor{LightGreen}{rgb}{0.88, 1, 0.88}
\newcommand{\cclb}{\cellcolor{LightBlue}}
\newcommand{\cclr}{\cellcolor{LightRed}}
\newcommand{\cclg}{\cellcolor{LightGreen}}

\allowdisplaybreaks

\def\gtwid{\mathrel{\raise.3ex\hbox{$>$\kern-.75em\lower1ex\hbox{$\sim
$}}}}
\def\vio{\mathrel{\hbox{$E$\kern-.60em\hbox{$/
$}}}}
\input macros.tex

\restylefloat{figure}

\begin{document}

\title{\bf Road map through the desert: unification with vector-like fermions}

\author{\\ Kamila Kowalska\let\thefootnote\relax\footnote{\url{kamila.kowalska@ncbj.gov.pl}}
and Dinesh Kumar\let\thefootnote\relax\footnote{\url{dinesh.kumar@ncbj.gov.pl}} 
\\[5ex]
\small {\em National Centre for Nuclear Research}\\
\small {\em Pasteura 7, 02-093 Warsaw, Poland }\\
}
%
\date{}
\maketitle
\thispagestyle{fancy}
\begin{abstract}
In light of null results from  New Physics searches at the LHC, we look at unification of the gauge couplings as a model-building principle. As a first step, we consider extensions of the Standard Model with vector-like fermions. We present a comprehensive list of spectra that feature fermions in two distinct $SU(3)_C\times SU(2)_L\times U(1)_Y$ representations, in which precise gauge coupling unification is achieved. We derive upper and lower limits on vector-like masses from proton decay measurements, running of the strong gauge coupling, heavy stable charged particle searches, and electroweak precision tests. We demonstrate that due to a particular hierarchy among the mass parameters required by the unification condition, complementarity of various experimental strategies allows us to probe many of the successful scenarios up to at least 10\tev.
\end{abstract}
\newpage 
%

\setcounter{footnote}{0}

\section{Introduction}\label{intro}

Unification of three fundamental forces of the Standard Model (SM) into a single gauge interaction has been an enticing idea since the mid 1970s\cite{Georgi:1974sy,Pati:1974yy,Mohapatra:1974hk,Fritzsch:1974nn,Georgi:1974my}. It emerged as a natural continuation of intellectual efforts that, in merging apparently unrelated phenomena, sought the key to a deeper understanding of nature, first by combining electricity and magnetism into a unified description, later leading to the establishment of the electroweak theory. Although the concept of unification as an underlying organizing principle stems to some extent from a sense of aesthetics,
it finds a more robust justification in the fact that the renormalized gauge couplings of the SM, while evaluated at higher and higher energies, seem to converge towards a common value. This behavior might be understood as a manifestation of a new, unified description of fundamental interactions known as Grand Unified Theory (GUT).

Precise gauge coupling unification, however, is not really achieved in the SM as discrepancies among the GUT-scale values of the SM couplings reach several percent. To make it work, the particle spectrum needs to be extended in order to modify the renormalization group (RG) running of the couplings below the GUT scale.\footnote{In principle, the GUT-scale values of the couplings can also be modified by high-scale threshold corrections\cite{Cook:1979pr,Dixit:1989ff,Langacker:1992rq,Hagiwara:1992ys}. These corrections, however, are strongly model dependent and for a certain range of the  GUT-particle masses they become negligibly small\cite{Schwichtenberg:2018cka}. Therefore, we neglect the effects of GUT threshold corrections throughout this study.} Supersymmetry (SUSY) has the advantage of leading to gauge unification in a quite natural way, yet no experimental evidence of the low-scale SUSY has been found so far. While this fact  does not undercut it completely as a theoretical framework, it is timely to ask to what extent unification of the gauge couplings is a unique property among various extensions of the SM. In other words, how many different beyond-the-SM (BSM) scenarios can be found whose particle spectrum differ quantitatively from the one of SUSY, and still allow for precise unification. 

In addressing this question we would like to remain as generic as possible. On the other hand, a truly comprehensive study of all imaginable SM extensions would be a highly challenging (if not impossible) task. For that reason our approach will be incremental: we are going to begin with a relatively simple BSM setup, which will then be gradually extended to encompass more complex structures. It is in this spirit that we regard the issue of unification as a long term research project, a {\it road map} that would guide the model building through the desert between the electroweak (EW) and GUT scales. 

We begin with defining the common framework for any unification analysis that we are going to undertake. The most important requirement is that the SM gauge symmetry persists up to the unification scale. It means, we will not consider Pati-Salam\cite{Pati:1974yy} or trinification\cite{Achiman:1979,Rujula:1984} type of GUTs as they do not require simultaneous unification of all three SM gauge couplings. For theoretical consistency, we also demand perturbativity of the renormalized model parameters up to the GUT energies.

Now we are going to make several additional assumptions, which on the one hand will substantially simplify the analysis, on the other will restrict the types of BSM scenarios that will be considered. Therefore, such assumptions may be dropped in the future studies. The additional requirements we impose are the following:
(i) any extension of the SM must be anomaly free; (ii) scenarios with low scale unification, $M_{\textrm{GUT}}\lsim 10^{15}\gev$, are not allowed (i.e.~general dimension-6 operators leading to proton decay are not forbidden by any additional mechanism\cite{Perez:2014kfa}). Since the first condition restricts possible BSM particles to vector-like (VL) fermions and scalars, for the sole purpose of the current study (iii) we will only consider VL extensions of the SM.

The issue of gauge coupling unification in the presence of VL fermions is not a new idea and it received a lot of attention in the literarture, both before the launch of the LHC\cite{Rizzo:1991tc,Zhang:2000zy,Choudhury:2001hs,Li:2003zh,Morrissey:2003sc,Giudice:2004tc,Dorsner:2005fq,EmmanuelCosta:2005nh,Shrock:2008sb,Gogoladze:2010in}, and after\cite{Dermisek:2012as,Dermisek:2012ke,Dorsner:2014wva,Xiao:2014kba,Bhattacherjee:2017cxh,Schwichtenberg:2018cka}. Similar analyses within the SUSY framework were performed as well\cite{ArkaniHamed:2004fb,Barger:2006fm,Barger:2007qb,Calibbi:2009cp,Donkin:2010ta,Liu:2012qua,Zheng:2017aaa,Zheng:2019kqu}.
In most of the former studies only particular types and a limited number of VL representations were considered. In this regard Ref.\cite{Rizzo:1991tc} took a more generic approach and looked at 84 different $SU(3)_C\times SU(2)_L\times U(1)_Y$ charge assignements for BSM fermions and scalars. Relatively recently the first attempt has been made in Ref.\cite{Bhattacherjee:2017cxh} to systematically study all possible VL extensions of the SM, in which BSM matter multiplets form incomplete representations of $SU(5)$. Scenarios with two distinct representations (and no more than six VL pairs in each of them) were considered, while independent VL masses were limited to 5\tev.\footnote{Results for 3- and 4-representation scenarios and the VL mass fixed at $1\tev$ were  shown in Ref.\cite{Bhattacherjee:2017cxh} as well. However, as we will demonstrate in the present study, mass hierarchy among various VL representations is one of the main factors of the successful gauge coupling unification. For this reason, the fixed-mass analyses can not be considered comprehensive.}

In the present work, we build on the findings of Ref.\cite{Bhattacherjee:2017cxh} and extend their analysis in several different directions. First of all, we boost the allowed mass range of VL fermions up to 10\tev. While it may seem far beyond the reach of modern colliders, we will show that due to a particular hierarchy among the VL spectra allowing for unification, as well as to complementarity of various experimental strategies, one is actually able to derive exclusion lower bounds even on multi-TeV masses. 
Secondly, we do not {\it a priori} limit the maximum number of VL pairs in each representation. It turns out that, when this extra condition is discarded, novel solutions with respect to Ref.\cite{Bhattacherjee:2017cxh} can be found. Finally, we thoroughly discuss a variety of experimental methods that allow one to test the successful unification scenarios. We derive upper and lower limits on VL masses from proton decay measurements, running of the strong gauge coupling, heavy stable charged particle searches, and EW precision tests. We demonstrate that by combining independent experimental results we manage in many cases to probe (and to exclude) essentially the whole parameter space of a given model. 

The paper is organized as follows. In \refsec{bsmvl} we define the fundamental building blocks of our BSM scenarios in terms of the transformation properties of VL fermions under the SM gauge symmetry group. \refsec{Analysis} presents the main results of the study: a comprehensive list of representations that allow for unification of the SM gauge couplings. In \refsec{bounds} we discuss in detail experimental bounds that constrain the parameter space of the successful scenarios. We present our conclusion in \refsec{conc}. Technical details of the analysis are collected in two appendices.

\section{Generic BSM scenarios with VL fermions}\label{bsmvl}

We begin our discussion with constructing a set of generic extensions of the SM that satisfy the requirements defined in the introduction, i.e. no extra gauge symmetry is imposed, the only new particles in the spectrum are VL fermions, and the renormalized couplings remain perturbative at the unification scale. We additionally assume for the purpose of this study that any Yukawa interactions generated by the BSM sector and allowed by the gauge symmetry are negligible.\footnote{The impact of non-gauge interactions on unification of the gauge couplings will be discussed elsewhere.}

We thus introduce $N_{F_i}$ copies of new fermionic fields, which transform under the $SU(3)_C\times SU(2)_L\times U(1)_Y$ gauge group as VL multiplets
\be
(R_{3F_i},R_{2F_i},Y_{F_i})\oplus(\bar{R}_{3F_i},\bar{R}_{2F_i},-Y_{F_i}).
\ee
Note that we count separately over both components of the pair, so $N_{F_i}$ can only assume even values (with an exception of fermions that transform in an adjoint representation of a non-abelian gauge symmetry group). The index $i$ runs over the number of distinct representations. At this stage both $i$ and $N_{F_i}$ are unconstrained. 

Upper bounds on the dimension of possible VL representations and on the number of fermions that transform accordingly are provided by perturbativity condition. Let us first consider the SM extended by one representation of VL fermions and assume $Y_F=0$. For various combinations of $R_3$ and $R_2$ and increasing number of VL copies, we run the SM gauge couplings from the low-energy scale, which we identify with the top quark mass $M_t$ and at which the couplings assume the following values\cite{Buttazzo:2013uya}
\be\label{init}
g_3(M_t)=1.16660\,,\qquad g_2(M_t)=0.64779\,,\qquad g_Y(M_t)=0.35830\,,
\ee
up to $10^{15}\gev$.\footnote{Note that we consider this particular value as a rough estimate of the limit imposed on the value of the GUT scale by proton decay measurement. The actual experimental bounds will be discussed in \refsec{bounds}} The VL mass is fixed at $10\tev$ as this is the largest allowed value of this parameter considered in the present study.  RG equations (RGEs) for the gauge couplings in a general quantum field theory are well known\cite{Machacek:1983tz} and we summarize their explicit two-loop form in \refap{rges}. 

\begin{table}[t]
\parbox{.45\linewidth}{
\centering
\begin{tabular}{|c|cccc|}
\hline
\diagbox{$\bm{R_{2}}$}{$\bm{R_{3}}$} & $\bm{1}$ & $\bm{3}$ & $\bm{6}$ & $\bm{8}$ \\
\hline
$\bm{1}$ & $\infty$ & 24 & 4 & 4 \\
$\bm{2}$ & 28 & 8 & 2 & 2 \\
$\bm{3}$ & 6 & 2 & 0 & 0 \\
$\bm{4}$ & 2 & 0 & 0 & 0\\
\hline
\end{tabular}
\caption{Maximal number of VL fremions with the mass of $10\tev$, which allows for perturbative gauge couplings below  $10^{15}\gev$. $Y_F=0$ is assumed.}\label{max_num_fer}
}
\hfill
\parbox{.45\linewidth}{
\centering
\begin{tabular}{|c|cccc|}
\hline
\diagbox{$\bm{R_{2}}$}{$\bm{R_{3}}$} & $\bm{1}$ & $\bm{3}$ & $\bm{6}$ & $\bm{8}$ \\
\hline
$\bm{1}$ & $3\frac{1}{6}$ & $1\frac{5}{6}$ & $1\frac{1}{6}$ & $1\frac{1}{6}$ \\
$\bm{2}$ & $2\frac{1}{3}$ & $1\frac{1}{3}$ & $\frac{5}{6}$ & $\frac{2}{3}$ \\
$\bm{3}$ & $1\frac{5}{6}$ & 1 & 0 & 0 \\
$\bm{4}$ & $1\frac{2}{3}$ & 0 & 0 & 0 \\
\hline
\end{tabular}
\caption{Maximal value of the hypercharge, which allows for perturbative gauge couplings below $10^{15}\gev$ ($N_F=2$ and the VL mass is set at $10\tev$).}\label{max_val_hyp}
}
\end{table}

In \reftable{max_num_fer} we show the maximal allowed number of VL fermions, $N_{F_{\textrm{max}}}$,  for which the gauge couplings remain perturbative ($g_i\lesssim 4\pi$) up to $10^{15}\gev$. One can see that color octets and electroweak quadruplets are the highest representations possible, and that a total number of 11 different combinations of $SU(3)_C$ and $SU(2)_L$ charges is allowed. Non-zero hypercharge can only reduce the value of $N_{F_{\textrm{max}}}$. 
Thus, the requirement of perturbativity up to around the unification scale reduces the possible number of VL fermions that transform under $SU(3)_C$ and $SU(2)_L$. In some cases, however, $N_{F_{\textrm{max}}}$ exceeds the maximum number of 12 VL copies adpoted in Ref.\cite{Bhattacherjee:2017cxh}. We will demonstrate in \refsec{Analysis} that several novel solutions with respect to those presented in Ref.\cite{Bhattacherjee:2017cxh} can be found if that somewhat arbitrary assumption is relaxed.

So far our discussion was quite generic as the conclusions regarding the properties of the allowed VL representations resulted merely from the requirement of perturbativity, independently on what happens at the unification scale and how the expected GUT symmetry is realized. It is, however, not the case for the hypercharge. 
This particular quantum number is much more difficult to deal with in a general manner, as in principle it can assume continuous values. Additionally, hypercharge normalization is not unique as it depends on a particular embedding of the SM into a GUT gauge group\cite{PerezLorenzana:1999tf}, and different normalizations may lead to different predictions regarding the gauge coupling unification. For these reasons we have to depart at this point from an entirely model-independent approach.

We assume from now on that at the unification scale the $SU(5)$ symmetry is restored and  VL fermions are embedded into multiplets of $SU(5)$ just like it is  the case for the SM fields. This seems to be the most natural choice since $SU(5)$ not only can play the role of a self-contained unified gauge symmetry\cite{Georgi:1974sy}, but also shows up in breaking chains of larger GUT groups. In \refap{su5} the decomposition of the irreducible $SU(5)$ representations of increasing dimensions into irreducible representations of the SM gauge group is summarized. It is enough to consider representations up to dimension 75, since the larger ones decompose either to representations that have already appeared in \refeq{su5reps}, or to representations whose dimensions  exceed the limits presented in \reftable{max_num_fer}. Additionally, in \reftable{max_val_hyp} we provide information about the maximal value of hypercharge for a single pair of VL fermions ($N_{F}=2$), for which the gauge couplings remain perturbative up to $10^{15}\gev$. When combined, perturbativity bounds in \reftables{max_num_fer}{max_val_hyp} eliminate some of representations listed in \refeq{su5reps}. Eventually, we are left with a set of 24 distinct non-singlet $SU(3)_C\times SU(2)_L\times U(1)_Y$ representations:
\bea\label{modlist}
\textrm{color singlets}: && \left({\bf 1}, {\bf 1}, 1\right), \left({\bf 1}, {\bf 1}, -2\right), \left({\bf 1}, {\bf 2}, \textstyle\frac{1}{2}\right), \left({\bf 1}, {\bf 2}, -\textstyle\frac{3}{2}\right), \left({\bf 1}, {\bf 3}, 0\right), \left({\bf 1}, {\bf 3}, 1\right), \\
&& \left({\bf 1}, {\bf 4}, \textstyle\frac{1}{2}\right), \left({\bf 1}, {\bf 4}, -\textstyle\frac{3}{2}\right),\nonumber\\
\textrm{color triplets}: && \left({\bf 3}, {\bf 1}, -\textstyle\frac{1}{3}\right), \left({\bf\bar{3}}, {\bf 1}, -\textstyle\frac{2}{3}\right), \left({\bf\bar 3}, {\bf 1}, \textstyle\frac{4}{3}\right), \left({\bf\bar 3}, {\bf 1}, -\textstyle\frac{5}{3}\right),  \left({\bf 3}, {\bf 2}, \textstyle\frac{1}{6}\right), \left({\bf\bar 3} , {\bf 2}, \textstyle\frac{5}{6}\right),  \nonumber\\
&&\left({\bf\bar 3}, {\bf 2}, -\textstyle\frac{7}{6}\right), \left({\bf 3}, {\bf 3}, -\textstyle\frac{1}{3}\right), \left({\bf\bar 3}, {\bf 3}, -\textstyle\frac{2}{3}\right), \nonumber\\
\textrm{color sextets}: && \left({\bf\bar 6}, {\bf 1}, -\textstyle\frac{1}{3}\right), \left({\bf 6}, {\bf 1}, -\textstyle\frac{2}{3}\right),  \left({\bf\bar 6}, {\bf 2}, \textstyle\frac{1}{6}\right), \left({\bf 6}, {\bf 2}, \textstyle\frac{5}{6}\right), \nonumber\\
\textrm{color octets}: && \left(\bm{8}, \bm{1}, 0\right), \left(\bm{8}, \bm{1}, 1\right), \left(\bm{8}, \bm{2}, \textstyle\frac{1}{2}\right). \nonumber
\eea
These are the fundamental building blocks of the VL unification scenarios we are going to analyze in the next section.

\section{VL models with precise gauge unification}\label{Analysis}

We are now in the position to perform a comprehensive analysis of the SM extensions with VL fermions that could potentially lead to precise unification of three SM gauge couplings at the energies in the range $[10^{15}-10^{18}]\gev$. With this goal in mind, we would like to proceed in a systematic way, gradually increasing the complexitity of the constructed models. The simplest scenarios could be engineered by adding to the SM one of the VL representations listed in \refeq{modlist}. It is, however, a known fact\cite{Bhattacherjee:2017cxh} that precise unification is not possible within such a framework. The next possibility is then to consider two different VL representations with an arbitrary number of fermion copies within each of them. Theoretical and phenomenological properties of such models are the main objective of the present study. 

Adding three (or more) independent VL representations makes our task more and more challenging. Note that 276 distinctive combinations of the VL representations listed in \refeq{modlist} need to be considered in the two-representation case. This figure increases to 2024 when three, and to 10626 when four different representations are considered. Each combination requires to scan over the numbers of VL fermions in each representation (see \reftable{max_num_fer}), as well as on their masses, which we always assume to be uncorrelated. This means that numerical complexity of the problem grows exponentially with every independent representation added. For this reason we focus in this study on the simplest case, leaving more complicated SM extensions for future work. 

The numerical procedure employed in our analysis is the following. We use the 2-loop SM RGEs from $M_t$ up to the scale $M_{1}$, at which the lightest of VL fermions show up in the spectrum. We assume for simplicity that all $N_F$ copies of the same representation have a common mass.  Above $M_{1}$ we switch to the 2-loop RGEs for a generic BSM scenario, \refeqst{beta1}{beta3}.
At the scale $M_2$ the effects of heavier VL fermions need to be taken into account. Finally, we define a unification scale, \mgut, as the scale at which all three gauge couplings acquire a common value, $g_{\gut}$,
\be
g_{\gut}=g_3(\mgut)=g_2(\mgut)=g_1(\mgut)\,.
\ee
We require that the unified coupling is perturbative, i.e. $g_{\gut}\leq 4\pi$.

Precision of the gauge coupling unification can be quantified by a set of three mismatch parameters, $\epsilon_{1,2,3}$.
For each combinations $(i,j)$, where $i,j=1,2,3$, we determine a two coupling unification scale from the condition $g_i(\mgut^{ij})=g_j(\mgut^{ij})=g_{ij}$. We then define a deviation of the third coupling from $g_{ij}$ as
\be
\epsilon_{k}=\frac{g_k^2(\mgut^{ij})-g^2_{ij}}{g^2_{ij}}
\ee
and determine the true unification scale by requiring $\epsilon_{\textrm{GUT}}=\min(\epsilon_1,\epsilon_2,\epsilon_3)$. In the SM  $\epsilon_{\textrm{GUT}}^{\textrm{SM}}=7.3\%$, indicating that the running values of the three gauge couplings do not really converge to a common number. 
On the other hand, in the minimal SUSY version of the SM, which is a benchmark BSM scenario for the gauge coupling unification, $\epsilon_{\textrm{GUT}}^{\textrm{MSSM}} = 1.1\%$ when all sparticle masses set at 1 TeV. 
Therefore, we define the {\it precise gauge unification} (PGU) by a condition $\epsilon < 1\%$.

\begin{table}[t]\footnotesize
\centering
\begin{tabular}{|c|cc|cc|c|}
\hline
Scenario & $\bm{R_{F_1}}$ & $\bm{R_{F_2}}$ & $N_1$ & $N_2$  &  VL mass/GUT scale \\
\hline
F1 & $\left(\bm{1}, \bm{2}, \frac{1}{2}\right)$ & $\left({\bf 6}, \bm{1}, \frac{1}{3}\right)$ & 12 & 2 &\reffig{fig:models1}(a) \\
F2 & $\left(\bm{1}, \bm{2}, \frac{1}{2}\right)$ & $\left({\bf 6}, \bm{1}, \frac{1}{3}\right)$ & 20 & 4 &\reffig{fig:models1}(b) \\
F3 & $\left(\bm{1}, \bm{2}, \frac{1}{2}\right)$ & $\left({\bf 6}, \bm{1}, \frac{1}{3}\right)$ & 22 & 4 &\reffig{fig:models1}(c) \\
F4 & $\left(\bm{1}, \bm{2}, \frac{1}{2}\right)$ & $\left(\bm{8}, \bm{1}, 0\right)$ & 8 & 1 &\reffig{fig:models1}(d) \\
F5 & $\left(\bm{1}, \bm{2}, \frac{1}{2}\right)$ & $\left(\bm{8}, \bm{1}, 0\right)$ & 12 & 2 &\reffig{fig:models1}(e) \\
F6 & $\left(\bm{1}, \bm{2}, \frac{1}{2}\right)$ & $\left(\bm{8}, \bm{1}, 0\right)$ & 14 & 2 &\reffig{fig:models1}(f) \\
F7 & $\left(\bm{1}, \bm{3}, 0\right)$ & $\left(\bm{3}, \bm{1}, -\frac{1}{3}\right)$ & 2 & 8 &\reffig{fig:models1}(g) \\
F8 & $\left(\bm{1}, \bm{3}, 0\right)$ & $\left(\bm{3}, \bm{1}, -\frac{1}{3}\right)$ & 3 & 12 &\reffig{fig:models1}(h) \\
F9 & $\left(\bm{1}, \bm{3}, 0\right)$ &  $\left(\bm{6}, \bm{1}, -\frac{2}{3}\right)$ & 3 & 2 &\reffig{fig:models1}(i) \\
F10 & $\left(\bm{1}, \bm{4}, \frac{1}{2}\right)$ &  $\left(\bm{6}, \bm{1}, -\frac{2}{3}\right)$ & 2 & 4 &\reffig{fig:models2}(a)\\
F11 & $\left(\bm{3}, \bm{1}, -\frac{1}{3}\right)$ &  $\left(\bm{3}, \bm{2}, \frac{1}{6}\right)$ & 2 & 2 & \reffig{fig:models2}(b)\\
F12 & $\left({\bf 3}, \bm{1}, \frac{2}{3}\right)$ &  $\left(\bm{3}, \bm{2}, \frac{1}{6}\right)$ & 4 & 4 & \reffig{fig:models2}(c)\\
F13 & $\left({\bf 3}, \bm{1}, \frac{2}{3}\right)$ &  $\left(\bm{3}, \bm{2}, \frac{1}{6}\right)$ & 6 & 6 &\reffig{fig:models2}(d)\\
\hline
\end{tabular}
\caption{Scenarios with 2 representations of VL fermions that allow for the  PGU ($\epsilon_{\textrm{GUT}}\leq 1\%$) and the associated unification scale lies in the range $10^{15}-10^{18}\gev$. The VL masses vary between 0.25\tev\ and 10\tev. In columns 2 and 3 transformation properties of both representations with respect to the SM gauge symmetry group are given. It is understood that, if applicable, $\bm{R_F}$ also encompasses its own complex conjugation. Columns 4 and 5 display number of VL fermions in each representation. In the last column we direct the reader to a corresponding figure illustrating the allowed VL mass ranges and the iso-contours of the unification scale. }
\label{2reps_all_comb}
\end{table}

For each of 276 scenarios with two distinctive VL representations we scan over the number of BSM fermions, $N_1,\,N_2$, and their masses, $M_1,\,M_2$, and for each point in the 4-dimensional parameter space we determine $\epsilon_{\textrm{GUT}}$ and $\mgut$. The parameters $M_1$ and $M_2$ are varied between 0.25\tev\ and 10\tev. The main results of the analyses are summarized in \reftable{2reps_all_comb}. We found 13 different scenarios that allow for the PGU at the scale $10^{15}-10^{18}\gev$. Nine of them have been previously identified in\cite{Bhattacherjee:2017cxh}, while the scenarios F2, F3, F6 and F13 present completely novel solutions, characterized by either more than 12 copies of VL fermions in one of the representations, or VL masses larger than 5\tev. Note that some of the successful combinations of representations allow for various choices of fermion numbers. This is, for example, the case for the scenarios F1, F2 and F3, in which VL fields transforming as $\left(\bm{1}, \bm{2}, \frac{1}{2}\right)$ and $\left(\bm{1}, {\bf\bar 2}, -\frac{1}{2}\right)$ of $SU(3)_C\times SU(2)_L\times U(1)_Y$ can show up in 12, 20 and 22 copies, while those transforming as $\left({\bf 6}, \bm{1}, \frac{1}{3}\right)$ and $\left({\bf\bar 6}, \bm{1}, -\frac{1}{3}\right)$ in 2 and 4 copies.  Thus, only 7 combinations of $R_3$ and $R_2$ are really unique.  

Among 24 various representations listed in \refeq{modlist}, only 9 can contribute to the successful unification. These are adjoint representations of both $SU(3)_C$ and $SU(2)_L$ gauge groups, $\left(\bm{8}, \bm{1}, 0\right)$ and $\left(\bm{1}, \bm{3}, 0\right)$; fundamental representations under which the SM left-handed quarks and leptons transform, $\left(\bm{3}, \bm{2}, \frac{1}{6}\right)$ and $\left(\bm{1}, \bm{2}, \frac{1}{2}\right)$; fundamental representations of the SM right-handed up and down quarks, $\left(\bm{3}, \bm{1}, -\frac{2}{3}\right)$, $\left(\bm{3}, \bm{1}, -\frac{1}{3}\right)$; and three exotic representations that are not realized by the ordinary matter. Incidentally, the resemblance of the quantum numbers characterizing VL fermions that allow for the PGU to the ones of the SM particles can have important phenomenological consequences once the Yukawa-driven mixing with the SM fermions is allowed.

\begin{figure}[t]
\centering
\subfloat[F1]{%
\includegraphics[width=0.3\textwidth]{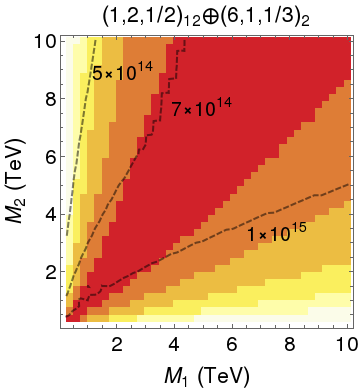}
}%
\hspace{0.1cm}
\subfloat[F2]{%
\includegraphics[width=0.3\textwidth]{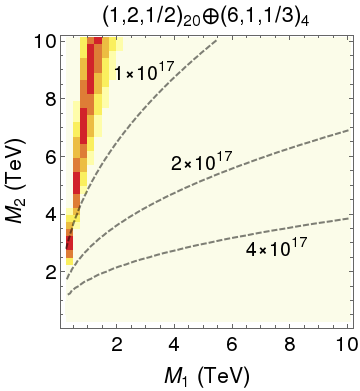}
}%
\hspace{0.1cm}
\subfloat[F3]{%
\includegraphics[width=0.3\textwidth]{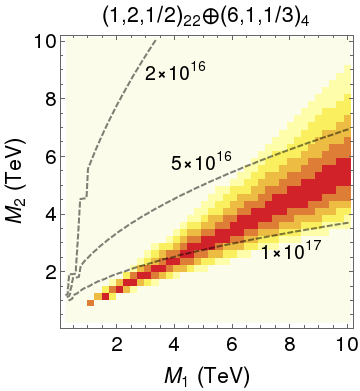}
}%
\hspace{0.0cm}
 \raisebox{0.5\height}{\includegraphics[width=0.03\textwidth]{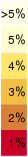}}
\\
\vspace{0.5cm}
\subfloat[F4]{%
\includegraphics[width=0.3\textwidth]{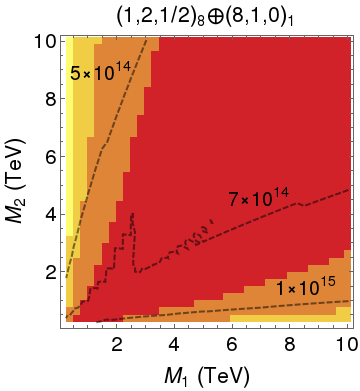}
}%
\hspace{0.1cm}
\subfloat[F5]{%
\includegraphics[width=0.3\textwidth]{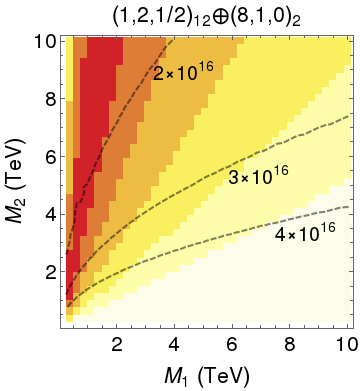}
}%
\hspace{0.1cm}
\subfloat[F6]{%
\includegraphics[width=0.3\textwidth]{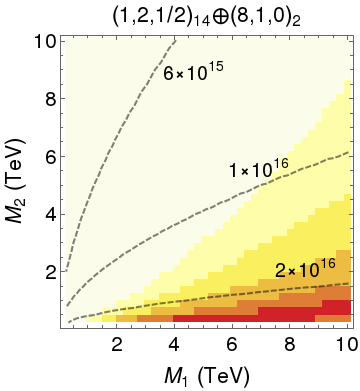}
}%
\hspace{0.0cm}
 \raisebox{0.5\height}{\includegraphics[width=0.03\textwidth]{Figs/legend.png}}
\\
\vspace{0.5cm}
\subfloat[F7]{%
\includegraphics[width=0.3\textwidth]{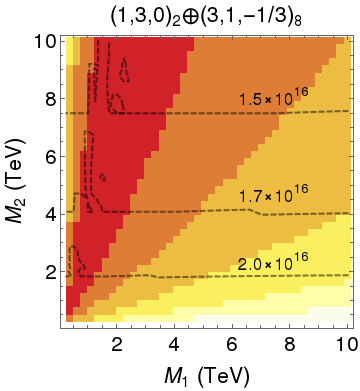}
}%
\hspace{0.1cm}
\subfloat[F8]{%
\includegraphics[width=0.3\textwidth]{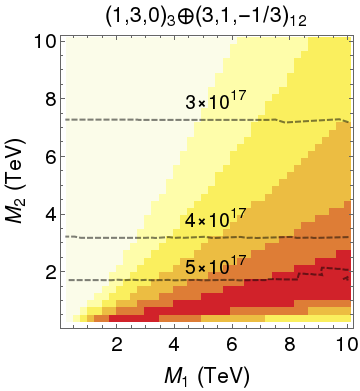}
}%
\hspace{0.1cm}
\subfloat[F9]{%
\includegraphics[width=0.3\textwidth]{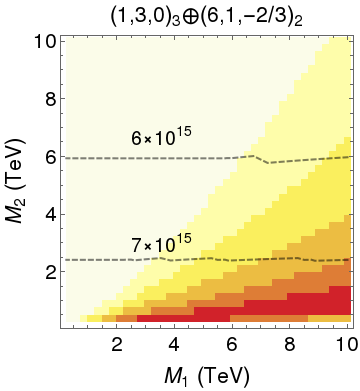}
}%
\hspace{0.0cm}
 \raisebox{0.5\height}{\includegraphics[width=0.03\textwidth]{Figs/legend.png}}
\caption{\footnotesize Distribution of the mismatch parameter $\epsilon_{\textrm{GUT}}$ as a function of $M_1$ and $M_2$ for the scenarios F1 - F9 of \reftable{2reps_all_comb}. Red area corresponds to the PGU ($\epsilon_{\textrm{GUT}}\leq 1\%$). Different shades of yellow illustrate departure from the precise unification condition as quantified by the increasing values of $\epsilon_{\textrm{GUT}}$. Isocontours of the unification scale (in \gev) are indicated as dashed black curves. 
\label{fig:models1}}
\end{figure}

\begin{figure}[t]
\centering
\subfloat[F10]{%
\includegraphics[width=0.3\textwidth]{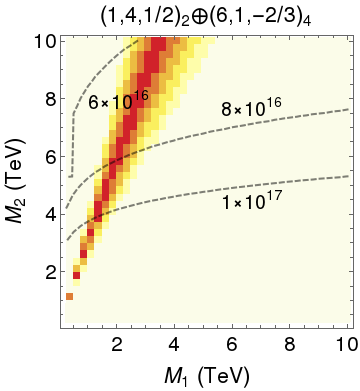}
}%
\hspace{0.1cm}
\subfloat[F11]{%
\includegraphics[width=0.3\textwidth]{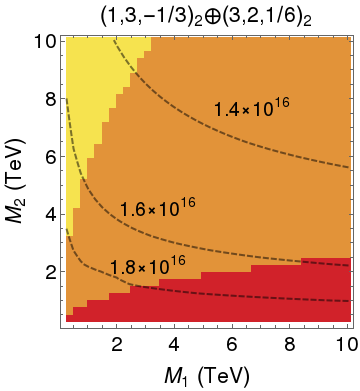}
}%
\hspace{0.0cm}
 \raisebox{0.5\height}{\includegraphics[width=0.03\textwidth]{Figs/legend.png}}
\\
\vspace{0.5cm}
\subfloat[F12]{%
\includegraphics[width=0.3\textwidth]{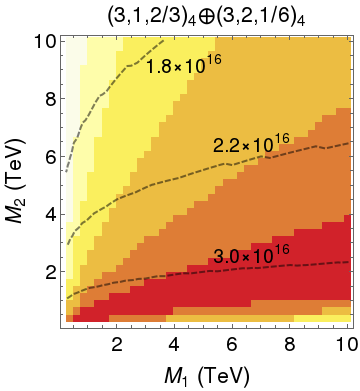}
}%
\subfloat[F13]{%
\includegraphics[width=0.3\textwidth]{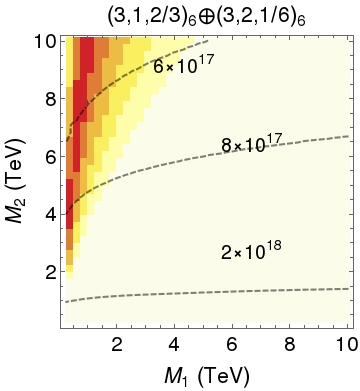}
}%
\hspace{0.0cm}
 \raisebox{0.5\height}{\includegraphics[width=0.03\textwidth]{Figs/legend.png}}
\caption{\footnotesize Distribution of the mismatch parameter $\epsilon_{\textrm{GUT}}$ as a function of $M_1$ and $M_2$ for the scenarios F10 - F13 of \reftable{2reps_all_comb}. Red area corresponds to the PGU ($\epsilon_{\textrm{GUT}}\leq 1\%$). Different shades of yellow illustrate departure from the precise unification condition as quantified by the increasing values of $\epsilon_{\textrm{GUT}}$. Isocontours of the unification scale (in \gev) are indicated as dashed black curves. 
\label{fig:models2}}
\end{figure}

It turns out that whether the gauge coupling unification is possible in a given model hinges strongly on hierarchy among the VL fermion masses. To illustrate this dependence, in \reffig{fig:models1} and \reffig{fig:models2} we present a distribution of the mismatch parameter $\epsilon_{\textrm{GUT}}$ as a function of $M_1$ and $M_2$ for all 13 scenarios summarized in \reftable{2reps_all_comb}. In red the region of the PGU is indicated, which will be of main interest for our further phenomenological analysis. As an additional information, we show in different shades of yellow departure from the precise unification condition as quantified by the increasing values of $\epsilon_{\textrm{GUT}}$. The width of the color bands can give one some idea on how easy the unification is, or, in other words, what is the required degree of fine-tuning among the mass parameters.
 
The successful PGU scenarios can be divided in three distinctive categories, depending on the required mass hierarchy among the VL fermions. We will be referring to them later using the following labels:
\bea
\rm{H0}:&& M_{1}\sim M_{2}\,,\qquad{\rm \;scenarios\;\; F1, F3, F4},\\
\rm{H1}:&& M_{1}\gg M_{2}\,,\qquad{\rm scenarios\;\; F6, F8, F9, F11, F12},\nonumber\\
\rm{H2}:&& M_{1}\ll M_{2}\,,\qquad{\rm scenarios\;\; F2, F5, F7, F10, F13}.\nonumber
\eea
In \refsec{bounds} we will demonstrate that the mass hierarchy characterizing a given scenario is crucial for the way the scenario can be tested experimentally.

Another quantity that significantly differentiates among the VL combinations listed in \reftable{2reps_all_comb} is the unification scale. In \reffig{fig:models1} and \reffig{fig:models2} its isocontours\ are indicated as dashed black curves. Depending on the order of magnitude of \mgut, we divide our scenarios in three categories:
\bea
{\rm Low}:&& \mgut\simeq 10^{15}\gev\,,\qquad{\rm scenarios\;\; F1, F4, F9},\\
{\rm Medium}:&& \mgut\sim 10^{16}\gev\,,\qquad{\rm scenarios\;\; F2, F5, F6, F7, F10, F11, F12},\nonumber\\
{\rm High}:&& \mgut\sim 10^{17}\gev\,,\qquad{\rm scenarios\;\; F3, F8, F13}.\nonumber
\eea

We summarize the characteristics of the PGU scenarios in terms of the VL mass hierarchy and the unification scale in \reftable{tab:class}. The color code refers to  experimental techniques that can be employed in order to test the available parameter space of a model. We will discuss them in details in \refsec{bounds}.

\begin{table}[t]\footnotesize
\centering
\begin{tabular}{|p{0.6cm}|C{1.7cm}|C{1.7cm}|C{1.7cm}|}
\hline
 & {Low} & {Medium} & {High} \\
\hline
& \cclb & & \\
\multirow{-2}{*}{H0} & \cclb\multirow{-2}{*}{\bf F1, F4}& & \multirow{-2}{*}{\bf F3}\\
\hline
& \cclg & \cclg{\bf F6, F11} & \cclg \\
\multirow{-2}{*}{H1} & \cclb\multirow{-2}{*}{\bf F9} & \cclg{\bf F12} & \cclg\multirow{-2}{*}{\bf F8}\\
\hline
 & \cclr & \cclr{\bf F2, F5}& \cclr\\
\multirow{-2}{*}{H2} &\cclb & \cclr{\bf F7, F10}& \cclr \multirow{-2}{*}{\bf F13}\\
\hline
\end{tabular}
\caption{\footnotesize Properties of the PGU scenarios in terms of the VL mass hierarchy and the unification scale and their susceptibility to various experimental search strategies. Light blue indicates the models that are tested by the proton decay measurements. Those highlighted in light green can be subject to color searches: $R$-hadrons and running of the strong gauge coupling. Light red corresponds to the scenarios tested through  EW interaction in lepton-like HSCP searches and EW precision tests.}
\label{tab:class}
\end{table}

Before closing this section, we would like to comment on the fate of the identified PGU scenarios when the VL masses are pushed to energies much higher than $10\tev$. To analyze this issue, we repeated the numerical procedure of \refsec{Analysis} extending the scanning ranges of $M_1$ and $M_2$ up to $10^{10}\tev$. We found that all the models listed in \reftable{2reps_all_comb}, except F1 and F4, remain valid at higher energies, as could be anticipated from the shape of the red areas in \reffig{fig:models1} and \reffig{fig:models2}. Additionally, several new combinations of two VL representations become available. 

The high energy behavior of the PGU scenarios may seem somehow discouraging, as in principle one may put the VL fermions well above the reach of any existing collider experiment and still achieve the gauge coupling unification. There is, however, one important remark to be made. The main factor that decides whether a given scenario is accepted or not, is the value of the unification scale, which we require to stay in the range $10^{15}-10^{18}\gev$. When the mass of VL fermions increases, the unification scale decreases, as confirmed by the shape of \mgut\ isocontours in \reffig{fig:models1} and \reffig{fig:models2}. As a consequence, experimental bounds from the proton decay measurements may at some point come into play. In the next section we will show that this is indeed the case and that the allowed parameter space of the PGU scenarios is limited from above.

\section{Experimental tests of the PGU scenarios}\label{bounds}

In the previous section we identified all possible combinations of two VL fermion representations that allow for precise unification of the three SM gauge couplings. In the following we will focus on phenomenological properties of the PGU scenarios and discuss in details various experimental ways of testing the available parameter space. We remind the reader that we assume negligible Yukawa couplings among the BSM sector and the SM, therefore our VL fermions can only be produced via gauge interactions. Experimental signatures of the models with a large number of VL fermions have been discussed by one of us in Ref.\cite{Bond:2017wut} and we follow closely its approach. We will demonstrate the complementarity among the bounds provided by various experimental searches, resulting from the fact that each of them aim at constraining particular sets of color and electroweak quantum numbers. In combination with the specific mass hierarchies required by the PGU (see \reftable{tab:class}), it will allow us to derive strong lower bounds on the VL masses.

\subsection{Proton decay}

We assume that at the unification scale the symmetry group $SU(3)_C\times SU(2)_L\times U(1)_Y$ is embedded into a larger GUT group. Since in the unified framework the SM quarks and leptons belong to the same GUT multiplets, interactions are generated, mediated by heavy gauge bosons, that violate both the baryon and lepton number conservation. Proton decay is then a generic prediction of such scenarios. In  non-SUSY models the dominant contribution to the proton decay width comes from dimension-6 gauge operators of the common structure $QQQL$. The exact form of these countributions highly depends on the realization of the GUT symmetry. However, a rough estimation of the proton lifetime can be made\cite{Nath:2006ut}
\be\label{life}
\tau_p=\left(\frac{4\pi}{g^2_{\textrm{GUT}}}\right)^2\left(\frac{\mgut}{\gev}\right)^4\times 2.0\times 10^{-32}\;\textrm{years},
\ee
as a simple function of the unification scale \mgut\ and the value of the unified gauge coupling $g_{\textrm{GUT}}$.

Proton decay has been experimentally searched for since the early 1990s by Super-Kamiokande (SK) underground water Cherenkov detector. The strongest lower bound on the proton lifetime is set by the decay channel $p\to e^+\pi^0$ and reads $\tau_p>1.6\times 10^{34}$ years\cite{Miura:2016krn}. Hyper-Kamiokande (HK), a next generation machine, will be able to extend the limit by at least one order of magnitude, up to $\sim 2\times 10^{35}$ years\cite{Abe:2018uyc}.

The present (solid blue line) and projected (dashed blue line) limits from the proton decay as a function of the VL masses $M_1$ and $M_2$ are shown in \reffigs{fig:bounds1}{fig:bounds2}. The shaded blue area above the lines is disfavored. Scenarios F1 and F4 are already entirely excluded by the SK measurements.
Scenario F9, on the other hand, is going to be entirely tested by HK. Scenarios that fall within the reach of the current proton decay experiments are also marked in \reftable{bounds} in light blue. As expected, all of them belong to the category ``low unification scale''.

Proton decay provides a unique experimental way of testing the PGU scenarios characterized by the BSM sector at the energy scales far above the reach of any present-day collider experiment. In fact, for all but one models from \reftable{2reps_all_comb} it provides upper bounds on the allowed VL masses. We report them in the second column of \reftable{exp_bounds} for the projected reach from HK (the corresponding current bounds from SK are approximately one order of magnitude weaker). One can see that for several scenarios that belong to the medium GUT-scale category the upper bounds on VL masses are of the order of ``only'' several-tens \tev. This feature opens up an exiting possibility of entirely probing those scenarios in the (however distant) future.

\begin{table}[t]\footnotesize
\centering
\begin{tabular}{|c|cc|cc|cc|cc|cc|c|}
\hline
& \multicolumn{2}{c|}{Proton decay}&\multicolumn{2}{c|}{Running $g_3$} &  \multicolumn{2}{c|}{$R$-hadrons} & \multicolumn{2}{c|}{HSCP}& \multicolumn{2}{c|}{EWPO}  & Summary \\
\multirow{-2}{*}{Model}&$M^{\textrm{max}}_1$ & $M^{\textrm{max}}_2$ &$M^{\textrm{min}}_1$ & $M^{\textrm{min}}_2$&$M^{\textrm{min}}_1$ & $M^{\textrm{min}}_2$&$M^{\textrm{min}}_1$ & $M^{\textrm{min}}_2$&$M^{\textrm{min}}_1$ & $M^{\textrm{min}}_2$ & plot\\
\hline
F1 &\multicolumn{2}{c|}{Excluded} & - & 0.7 & - & 1.8 &0.8 & -&1.7& - &\reffig{fig:bounds1}(a) \\
F2 & 25 & 180 & - & 1.1 & - & 1.8 &0.8&(6.0) & 2.0 & - &\reffig{fig:bounds1}(b) \\
F3 & 350 & 200 &  - & 1.1 & (2.2) & 1.8 &0.8&- &2.2 & - &\reffig{fig:bounds1}(c) \\
F4 & \multicolumn{2}{c|}{Excluded}& - & 0.4 & (1.2) & 2.0 &0.8&- & 1.2& -&\reffig{fig:bounds1}(d) \\
F5 & 10 & 50& - & 0.8 & - & 2.0 &0.8& (3.0)& 1.5& -&\reffig{fig:bounds1}(e) \\
F6 &500 & 50& (9.0) & 0.8 & ($>$10) & 2.0 &0.8&- & 1.7 & -& \reffig{fig:bounds1}(f) \\
F7 &20 & 100 & - & 0.5 & - &1.7 &1.1& -& 1.2& -&\reffig{fig:bounds1}(g) \\
F8 & $2\times10^{5}$&$5\times10^{5}$& (3.0) & 0.8 & (6.0) & 1.7 &1.1&- &1.2 & -& \reffig{fig:bounds1}(h) \\
F9 & \multicolumn{2}{c|}{Excluded HK} & (4.5) & 0.7 & ($>$10) & 1.8 &1.1&- &1.5 & -&\reffig{fig:bounds1}(i) \\
F10 &250 & 1000& - & 1.1 &- &1.8 &1.2 & (3.0) &2.0 & -&\reffig{fig:bounds2}(a)\\
F11 &600 & 200& 0.2 & 0.2 & (5.0) & 1.8 &-&- & 0.5&1.0&\reffig{fig:bounds2}(b)\\
F12 &$6\times10^{4}$& 400& 0.2 & 0.6 & (4.0) & 1.8 &-&- &0.7 &1.5&\reffig{fig:bounds2}(c)\\
F13 &- &$2\times10^{6}$& 0.5 & 0.8 & 1.7 & ($>$10) &-&- & 0.7& 1.7&\reffig{fig:bounds2}(d)\\
\hline
\end{tabular}
\caption{Exclusion bounds on the VL masses $M_1$ and $M_2$ provided by different experiments (all in \tev). In column 2 we indicate the models that are exluded by the measurement of the proton lifetime by Super-Kamiokande\cite{Miura:2016krn}, as well as the upper bounds provided by the projected Hyper-Kamiokande measurement\cite{Abe:2018uyc}. In column 3 limits from the running strong coupling constant measurement by CMS\cite{Khachatryan:2016mlc} are shown. The numbers in parentheses indicate indirect limits whose derivation is described in the text. In columns 4 and 5 bounds from the ATLAS 13 TeV HSCP searches\cite{Aaboud:2019trc} are presented, for colored and non-colored particles, respectively. 100 TeV projections for the EWP tests\cite{Farina:2016rws} are shown in column 6.}
\label{exp_bounds}
\end{table}

\subsection{Running of the strong gauge coupling}\label{runas}

The RG running of the strong gauge coupling constant has been tested experimentally up to the energies of around $1.5\tev$. 
The most recent data comes from the measurement of double-differential inclusive jet cross section at \eight\ with an integrated luminosity of $19.6\invfb$~by the CMS Collaboration\cite{Khachatryan:2016mlc}. The value of the running coupling is extracted from the data as a function of the energy scale at which it is evaluated. The measurement is consistent with the predictions of the SM and as such poses a constraint on the minimal mass of any exotic colored particle. 

\begin{figure}[t]
\centering
\subfloat[Scenario F1]{%
\includegraphics[width=0.3\textwidth]{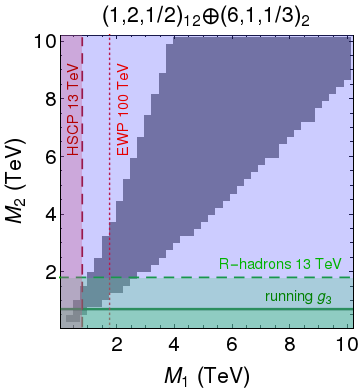}
}%
\hspace{0.1cm}
\subfloat[Scenario F2]{%
\includegraphics[width=0.3\textwidth]{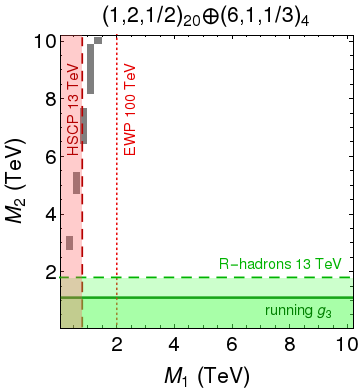}
}%
\hspace{0.1cm}
\subfloat[Scenario F3]{%
\includegraphics[width=0.3\textwidth]{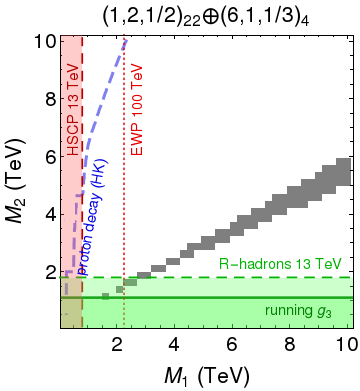}
}%
\\
\vspace{0.1cm}
\subfloat[Scenario F4]{%
\includegraphics[width=0.29\textwidth]{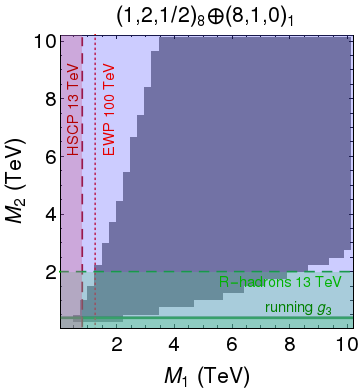}
}%
\hspace{0.1cm}
\subfloat[Scenario F5]{%
\includegraphics[width=0.29\textwidth]{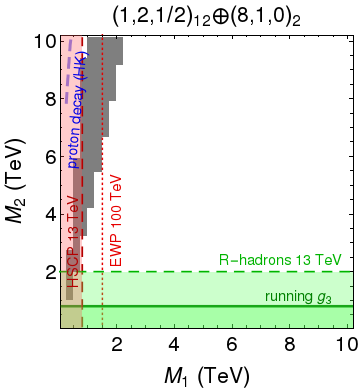}
}%
\hspace{0.1cm}
\subfloat[Scenario F6]{%
\includegraphics[width=0.29\textwidth]{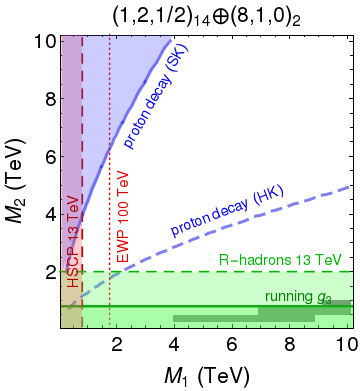}
}%
\\
\vspace{0.1cm}
\subfloat[Scenario F7]{%
\includegraphics[width=0.29\textwidth]{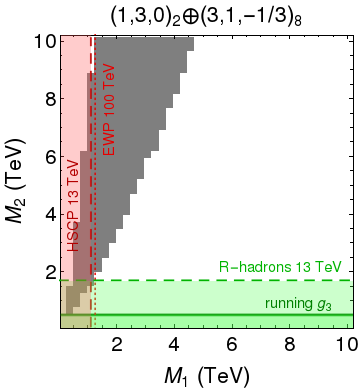}
}%
\hspace{0.1cm}
\subfloat[Scenario F8]{%
\includegraphics[width=0.29\textwidth]{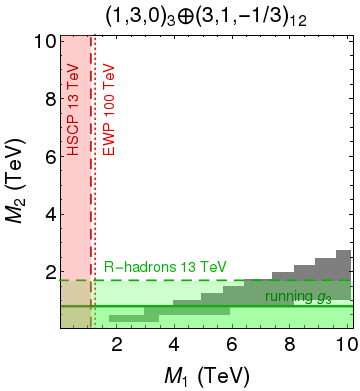}
}%
\hspace{0.1cm}
\subfloat[Scenario F9]{%
\includegraphics[width=0.29\textwidth]{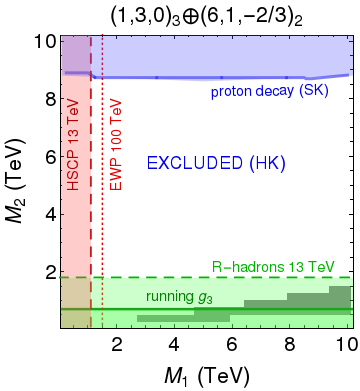}
}%
\caption{\footnotesize Summary of experimental bounds on VL masses $M_1$ and $M_2$ for scenarios F1-F9. In gray the PGU region is indicated. The area below and left to the solid green line is excluded by the measurement of the running strong coupling constant by CMS\cite{Khachatryan:2016mlc}. The limits from the 13 TeV ATLAS $R$-hadrons search\cite{Aaboud:2019trc} are indicated as a dashed green line. The corresponding lepton-like HSCP search excludes the area left to the dashed red line. 100 TeV projections for the EWP tests\cite{Farina:2016rws} are depicted as red dotted lines. Blue shaded region marks the exlusion by the proton decay measurement at Super-Kamiokande\cite{Miura:2016krn}. A projected reach of Hyper-Kamiokande\cite{Abe:2018uyc} is shown as a blue dashed line. \footnotesize 
\label{fig:bounds1}}
\end{figure}

In the third column of \reftable{exp_bounds} we summarize the lower bounds on the VL masses $M_1$ and $M_2$ in each PGU scenario. The same limits are also depicted in \reffigs{fig:bounds1}{fig:bounds2} as dark green solid lines. Obviously, only the representations that transform non-trivially under $SU(3)_C$ can be directly constrained by the data.

There is, however, an interesting observation to be made. In the scenarios with the mass hierarchy H1, characterized by the color VL fermions much lighter than the non-colored ones, the running strong coupling allows one to indirectly put very strong lower bounds on masses of the fermions that are $SU(3)_C$ singlets and would be otherwise not affected by the CMS measurement.  We indicate them in \reftable{exp_bounds} as numbers in parentheses.  As an example, let us consider scenario F6, whose parameter space is subject to various experimental constraints presented in \reffig{fig:bounds1}(f). The direct lower bound on $M_2$ from the running of $g_3$ reads in this case $0.8\tev$. The PGU region, however, is located in the lower part of the plot, as the unification requirement imposes $M_1\gg M_2$. As a result, it is almost entirely probed by CMS and an indirect bound on the mass $M_1$ can be derived, which reads in this case $M_1\gsim 9\tev$.
We will later see that the indirect limits from the running of the strong coupling constant are actually stronger that any other bound  provided by dedicated electroweak searches. 

Enhanced susceptibility to color searches is marked in \reftable{bounds} in light green and, as explained above, it corresponds to the mass hierarchy H1. It is yet another example of the complementarity among particular properties of the PGU scenarios and their testability.

\begin{figure}[t]
\centering
\subfloat[Scenario F10]{%
\includegraphics[width=0.3\textwidth]{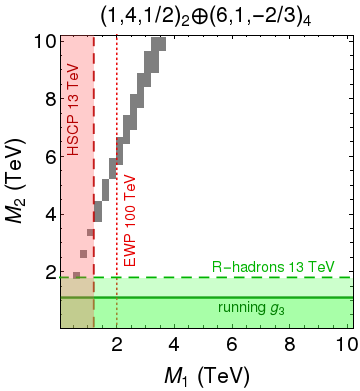}
}%
\hspace{0.1cm}
\subfloat[Scenario F11]{%
\includegraphics[width=0.3\textwidth]{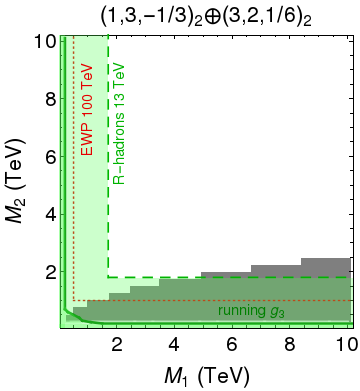}
}%
\\
\vspace{0.1cm}
\subfloat[Scenario F12]{%
\includegraphics[width=0.3\textwidth]{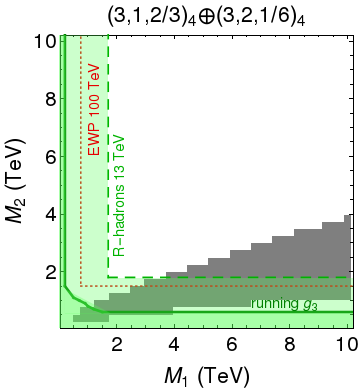}
}%
\subfloat[Scenario F13]{%
\includegraphics[width=0.3\textwidth]{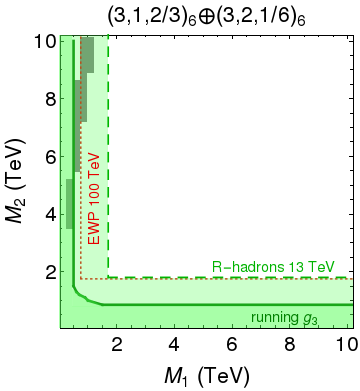}
}%
\caption{\footnotesize The same as \reffig{fig:bounds1} but for scenarios F10-F13.
\label{fig:bounds2}}
\end{figure}

\subsection{Direct LHC searches}

In the absence of Yukawa interactions with the SM quarks and leptons, the VL fermions are stable\footnote{The presence of a stable charged particle at cosmological scales may be problematic from the point of view of dark matter properties. A way out is to introduce Yukawa interaction with the SM, small enough not to affect the RG running but large enough to allow the charged particle to decay. Note, however, that in the case of representations $({\bf 8},{\bf 1},0)$, $({\bf 6},{\bf 1},\frac{1}{3})$ and $({\bf 6},{\bf 1},\frac{2}{3})$ it is not possible to construct a decay operator with the SM matter. One would need to introduce, for example, additional scalars charged under $SU(3)_C$. This provides a motivation to extend the current analyses in the future to include the scalar fields.} and can be experimentally looked for at  colliders through heavy stable charged particle (HSCP) searches. 
Dedicated analyses performed both by the ATLAS and CMS collaborations utilize observables related to the ionization energy loss ($dE/dx$) and time of flight (ToF), which allow to distinguish massive and non-relativistic HSCPs from the light SM particles traveling with velocities close to the speed of light.

Two categories of signals are usually considered, depending on the type of charges carried by HSCP: one that consists of particles interacting strongly, and another in which HSCPs are lepton-like color singlets. The two would differ both by the production mechanism at the LHC and by the size of the production cross section. In the following we will discuss them separately.

\subsubsection{Colored HSCP}

Let us first assume that a heavy stable particle can interact strongly. If the lifetime of such a colored  HSCP is longer than typical hadronization time scale, it can form colorless QCD bound states with the SM quarks and gluons, the so-called $R$-hadrons.

The most recent ToF and $dE/dx$ based analyses have been performed by ATLAS using a data sample corresponding to 36 fb$^{-1}$ of proton-proton collisions at $\sqrt{s}=13$ TeV\cite{Aaboud:2019trc}, and by CMS using 2.5 fb$^{-1}$ of data at the same energy\cite{Khachatryan:2016sfv}. Since in both cases no significant deviations from the expected SM background have been observed, a model-independent 95\% confidence level (C.L.) upper bound on the $R$-hadron production cross section can be derived. Such a result can then be translated into a lower bound on the BSM fermion mass within an arbitrary framework. An example usually considered by the collaborations is the gluino, the SUSY partner of the gluon and a benchmark for a BSM fermion with the SM charges $({\bf 8},{\bf 1},0)$. The lower bound on the long-lived gluino mass reads 1.5\tev\ for CMS  and 2.0\tev\ for ATLAS. 

In hadron colliders, any colored BSM fermion would be pair-produced at the leading order through gluon fusion or by quark-antiquark annihilation, with the  production cross section that solely depends on the $SU(3)_C$ quantum numbers. We calculated the $pp\to \bar{Q}Q$ cross section at the leading order (LO) using \madgr, and then rescaled it with the $k$-factor of $2$\cite{Tanabashi:2018oca} to account for higher-order QCD corrections and to reproduce the cross section quoted in\cite{Aaboud:2019trc} for gluino pair-production.\footnote{The cross section changes by over three orders of magnitude over the VL mass range considered in\cite{Aaboud:2019trc}. The resulting exclusion bound is, therefore, very mildly sensitive to higher-order order corrections to the cross section.} We then compared the result with the observed exclusion limit on the gluino derived by ATLAS. The corresponding exclusion bounds applied to parameters $M_1$ and $M_2$ are indicated in \reffigs{fig:bounds1}{fig:bounds2} as dashed green lines. We also summarize them in the fourth column of \reftable{exp_bounds}.

The limits from the $R$-hadron searches probe the parameter space in the same direction as the measurement of the running strong coupling constant from \refsec{runas}. Therefore, the mass parameter of the colored VL representations is constrained. As before, indirect bounds on the non-colored representations can be derived in the case of the type H1  mass hierarchy. The effect is particularly visible for scenarios F6 and F9, which turn out to be excluded by the $R$-hadron searches up to at least 10\tev. In several other scenarios $M_1$ receives a strong lower bound as well, which we indicate in \reftable{exp_bounds} as a number in  parentheses.

Note that in the presence of non-zero Yukawa interactions the VL colored fermions may decay before a bound state is formed. In such a case, the limits from the $R$-hadron searches will no longer apply. For that reason we indicate them in \reffigs{fig:bounds1}{fig:bounds2} with dashed lines, as contrasted with the running strong coupling constraints that are model-independent once the $SU(3)_C$ charges are fixed.

\subsubsection{Lepton-like HSCP}\label{lHSCP}

If a charged HSCP does not interact hadronically, it will be produced through Drell-Yan (DY) processes and will predominantly lose energy via ionization inside the detector. In the analysis\cite{Aaboud:2019trc} with 36 fb$^{-1}$ of data  ATLAS interpreted the model-independent results in a benchmark model that assummes DY production of charginos. The corresponding lower limits on the HSCP  mass reads 1090\gev. In the analogous study by CMS\cite{Khachatryan:2016sfv} based on 2.5 fb$^{-1}$ dataset, bounds on the mass of a generic lepton-like fermions with a unit electric charge were derived at $550\gev$.

To set lower bounds on the masses of VL fermions that are $SU(3)_C$ singlets, we used the chargino-dedicated search  by ATLAS. The LO production cross sections were calculated with \madgr, but no $k$-factor was added. The corresponding exclusion bounds in the  $(M_1,M_2)$ plane are depicted in \reffigs{fig:bounds1}{fig:bounds2} as dashed red lines. We also summarize them in the fifth column of \reftable{exp_bounds}.

The limits from the lepton-like HSCP searches allow to probe the scenarios with the mass hierarchy H2, in which the non-colored VL fermions are lighter than the colored ones. Enhanced susceptibility to electroweak searches is marked in \reftable{bounds} in light red.
In general, the limits are significantly weaker than the corresponding color-based bounds due to the lower production cross section. On the other hand, indirect lower bounds on the mass $M_2$ can be derived in scenarios F2, F5, F10, which turns out to be much stronger than the direct bounds from the $R$-hadron searches or the running strong coupling measurement. 
\subsection{Electroweak precision tests}

A complementary way to study properties of the VL fermions is to look at the processes below the \mvl\ mass threshold. Such an approach can result particularly important if VL fermions are too heavy to be directly produced in the colliders, or not long-lived enough for dedicated HSCP searches to be effective. In this regard, high-energy measurements of DY processes at the LHC offer a promising way to indirectly look for VL fermions by testing departures from the SM predictions in  electroweak precision (EWP) observables\cite{Farina:2016rws}.

In the VL extensions of the SM considered in this paper, the BSM contributions can manifest themselves in two oblique parameters\cite{Barbieri:2004qk,Cacciapaglia:2006pk} that are sensitive to the presence of states charged under the EW gauge symmetry, $W$ and $Y$.  The experimental bounds on $W$ and $Y$ are derived from the measuerements of charged and neutral currents DY  at hadron colliders. The VL fermion contributions to the parameters $W$ and $Y$ are directly related to the corresponding beta functions and given by\cite{Alves:2014cda}
\begin{align}\label{cwyone}
W,Y 
=\frac{g_{2,1}^2}{80\pi^2} \frac{m_W^2}{\mvl^2} \times \Delta B_{2,1}^{\textrm{BSM}} \, .
\end{align}
Here, $\Delta B_{2,1}^{\textrm{BSM}}$ denote the pure BSM contributions to the one-loop coefficients $B_{2,1}$, \refeq{oneloop2} and  \refeq{oneloop1}, respectively. 

\begin{table}[t]\footnotesize
\centering
\begin{tabular}{|c|c|c|}
\hline
Scenario & Current status & Experimental test\\
\hline
F1 & Excluded & proton decay\\
F2 & $M_1>0.8\gev$, $M_2>6.0\gev$ & HSCP\\
F3 & $M_1>2.2\gev$, $M_2>1.8\gev$ & $R$-hadrons\\
F4 & Excluded & proton decay\\
F5 & $M_1>0.8\gev$, $M_2>3.0\gev$ & HSCP\\
F6 & Excluded up to 10\tev & $R$-hadrons\\
F7 & $M_1>1.1\gev$, $M_2>1.7\gev$ & $R$-hadrons, HSCP\\
F8 & $M_1>6.0\gev$, $M_2>1.7\gev$ & $R$-hadrons\\
F9 & To be tested by HK &  proton decay\\
F10 & $M_1>1.2\gev$, $M_2>3.0\gev$ &  HSCP\\
F11 & $M_1>5.0\gev$, $M_2>1.8\gev$ & $R$-hadrons\\
F12 & $M_1>4.0\gev$, $M_2>1.8\gev$ & $R$-hadrons\\
F13 & Excluded up to 10\tev & $R$-hadrons\\
\hline
\end{tabular}
\caption{Summary of current experimental status of the successful PGU scenarios.}
\label{exp_summary}
\end{table}

The most up-to-date EWP experimental limits have been presented in\cite{Farina:2016rws}, including data from LEP\cite{Falkowski:2015krw} and LHC 8 TeV measurements by ATLAS\cite{Aad:2016zzw} and CMS\cite{CMS:2014jea}. We checked that they do not provide any bounds on the parameter space of the PGU scenarios under study.
However, since the effects of $W$ and $Y$ on DY processes grow with energy, the present experimental bounds can be significantly improved at the future colliders, by roughy two orders of magnitude at the projected 100 TeV machine\cite{Farina:2016rws}. The corresponding projections with 3 ab$^{-1}$ are depicted in \reffigs{fig:bounds1}{fig:bounds2} as dotted red lines. They are also summarized in the sixth column of \reftable{exp_bounds}. 

As expected from the size of the corresponding gauge couplings and group-theoretical factors, the constraints on $W$ are stronger than those on $Y$. Therefore, the projected EWP bounds are particularly powerful for VL representations with the non-trivial $SU(2)_L$ charges. As a consequence, in most cases it is the mass parameter $M_1$ that can be directly constrained by the EWP tests. The only exceptions are scenarios F11-F13, in which both VL representations can be constrained. Note also that in those three cases the projected EWP bounds can actually be competitive with the present day measurement of the running strong gauge coupling constant.
Finally, it is worth to stress that similarly to what we observed in \refsec{lHSCP} for the lepton-like HSCP searches, in the PGU scenarios with the mass hierarchy H2 indirect lower bounds on the mass of the $SU(2)_L$ singlet representations $M_2$ can be obtained.
\medskip

To summarize the findings of this section, we collect in \reftable{exp_summary} information about the current experimental status of the successful PGU scenarios. In this regards, we can divide them into three distinct categories. The first one encompasses scenarios F1 and F4, which are already excluded by the proton decay measurements, and scenario F9, which will be entirely tested by Hyper-Kamiokande. The second category corresponds to those scenarios (F6 and F13) that are excluded up to at least 10\tev, but which become allowed once higher VL masses are considered. The remaining eight scenarios feature the parameter space that still evades experimental bounds for VL masses in the multi-\tev\ regime. It should be noted, however, that some of them (F2, F5, F8 and F11) could be in the future and with more data entirely tested within the considered mass range by the HSCP searches, while two others (F7 and F12) can be tested for the most part. Scenarios F3, and F10, on the other hand, will remain more challenging to explore.

\section{Conclusions}\label{conc}

In light of null results from  New Physics searches at the LHC, we look at unification of the gauge couplings as a model-building principle and classify possible SM extensions that feature this property.

As a first step, we considered  in this study extensions of the SM with two distinct representations of VL fermions. We analyzed all their possible combinations with  the number of fermions in each representation limited only by perturbativity of the gauge couplings at the unification scale. We found 13 different combinations of two representations that allow for precise gauge unification at energies higher than $10^{15}\gev$, and for VL masses in the range $0.25-10\tev$. 

Interdependence between types of spectra required by the unification condition and their susceptibility to experimental tests is the main characteristics of successful PGU scenarios. 
We showed that the effectiveness of a given search in probing the allowed parameter space of a  model is directly related to its two features: mass hierarchy among VL fermions and the value of the unification scale.  Scenarios in which the colored fermions are much lighter than the non-colored ones may be almost entirely tested by the measurement of the running strong coupling and by the LHC $R$-hadron searches. And {\it vice versa}, if non-colored fermions are much lighter, HSCP searches and EW precision tests become very effective. On the other hand, scenarios in which both VL masses are of the same order remain beyond the reach of present-days colliders. In this case, however, null outcome from the proton decay experiments allows to exclude those models that feature the low unification scale.  

The results presented in this study clearly highlight the importance of combining different experimental strategies in order to derive the most robust constraints on the PGU parameter space. In this regard, proton decay measurements play a particular role, as they offer the only mean of probing the VL spectra above the multi-\tev\ regime. There is also a great potential in the direct HSCP searches at the LHC. We hope that our results will prove useful for experimental collaborations in choosing benchmark BSM scenarios for their future analyses.


The current study can be extended in different directions. First of all, one may consider more complex (and more realistic) BSM scenarios, featuring for example more than two VL representations or extra scalars. Secondly, the effects of non-gauge interactions (Yukawa and scalar types) should be taken into account, as they are bound to affect the phenomenology of PGU scenarios. After all, the desert seems like an interesting place to explore.

\bigskip\bigskip
\noindent \textbf{ACKNOWLEDGMENTS}
\medskip

\noindent 
We would like to thank Enrico Sessolo for his comments on the manuscript. The use of the CIS computer cluster at the National Centre for Nuclear Research in Warsaw is gratefully acknowledged. This work is supported by the National Science Centre (Poland) under the research Grant No.~2017/26/E/ST2/00470.
\appendix

\section{Group invariants and beta functions}\label{rges}

General two-loop beta functions for a system of gauge couplings $g_i$ of a direct-product symmetry group $G_i\times\cdots$ read\cite{Machacek:1983tz}
\bea\label{rgesall}
\beta_{i} =\frac{dg}{d\ln \mu}&=& \frac{g_i^3}{(4\pi)^2}\left[-\frac{11}{3}C_2(G_i)+\frac{2}{3}S_{2}(R_{Fi})+\frac{1}{3}S_{2}(R_{Si})\right]\\
&+&\frac{g_i^5}{(4\pi)^4}\left[-\frac{34}{3} C_2(G_i)^2+\Big(2C_2(R_{Fi})+\frac{10}{3}C_2(G_i)\Big)S_2(R_{Fi})+\Big(4C_2(R_{Si})+\frac{2}{3}C_2(G_i)\Big)S_2(R_{Si}) \right.\nonumber\\
&+&\left.\sum_{j=1}^kg_j^2\Big(2C_2(R_{Fj})S_2(R_{Fi})+4C_2(R_{Sj})S_2(R_{Si})\Big)\right],\nonumber
\eea
where $G_i$, $R_{Fi}$ and $R_{Si}$ denote contributions from gauge bosons, Weyl fermions, and complex scalars respectively. $C_2(R)$ is a quadratic Casimir invariant, $S_2(R)$ a Dynkin index of a representation $R$, and the sum is meant in both $S_2(R_{Fi})$ and $S_2(R_{Si})$ over all fermion and scalar representations transforming nontrivially under $G_i$. 

The quadratic Casimir operator for the representation $R$ of a symmetry group $G$ is defined as
\be
 C_2(R)\delta^i_j=(t^At^A)^i_j=\sum_{A=1}^d t^At^A,
\ee
where $t^A$ are the generators of $G$ in the representation $R$. The Dynkin index of a representation $R$ is instead given by
\be
 S_2(R)\delta^{A\,B}=\textrm{Tr}\left\{t^At^B\right\}.
\ee
The two are related through the dimensions of the representation $R$, $d(R)$, and of the adjoint, $d(\textrm{Adj})$,
\be
 S_2(R)d(\textrm{Adj})=C_2(R)d(R).
\ee
It is convenient to parameterize the quadratic Casimir operator, the Dynkin index, and the dimension of the representation through the weights $(p,q)$ for irreducible $SU(3)$ representations $R_3$, and, similarly, through the highest weight $\ell$ for  $SU(2)$ representations $R_2$, 
\be
\begin{array}{l}
d(R_3)=\frac12(p+1)(q+1)(p+q+2)\,,\\[1ex]
C_2(R_3)=p+q+\frac13 (p^2+q^2+p q)\,,\quad {\rm with}\quad p,q=0,1\cdots\,,\\[1ex]
d(R_2)=2\ell+1\,,\\[1ex]
C_2(R_2)=\ell(\ell+1) \,, \quad{\rm with}\quad  \ell=0,\frac{1}{2},1\cdots\,.
\end{array}
\ee

The two-loop beta functions for the SM augmented with $N_F$ fermions in the representation $(R_{F3},R_{F2},Y_F)$ are straightforwardly derived from\ref{rgesall}, and read
\bea
\label{beta1}\beta_3 &=& \frac{g_3^3}{(4\pi)^2} B_3+\frac{g_3^3}{(4\pi)^4}\left(C_{33}\, g_3^2  + C_{32}\, g_2^2 +C_{31}\, g_1^2\right),\\
\beta_2 &=& \frac{g_2^3}{(4\pi)^2} B_2+\frac{g_2^3}{(4\pi)^4}\left(C_{23}\, g_3^2  + C_{22}\, g_2^2 +C_{21}\, g_1^2\right),\\
\label{beta3}\beta_1 &=& \frac{g_1^3}{(4\pi)^2} B_1+\frac{g_1^3}{(4\pi)^4}\left(C_{13}\, g_3^2  + C_{12}\, g_2^2 +C_{11}\, g_1^2\right),
\eea
with the one-loop coefficients determined as
\bea
B_3 &=& -7+\frac{2}{3}N_F\,S_2(R_{F3})\,d(R_{F2}), \label{oneloop3}\\[1.5ex]
B_2 &=&-\frac{19}{6}+\frac{2}{3} N_F\, S_2(R_{F2})\,d(R_{F3}), \label{oneloop2}\\[1.5ex]
B_1 &=&\frac{41}{10}+\frac{2}{5} N_F\, d(R_{F3}) d(R_{F2}) Y_F^2.\label{oneloop1}
\eea
The two-loop coefficients are given by
\bea\label{twoloop}
C_{33}&=&-26+ N_F \,S_2(R_{F3})\,d(R_{F2})\big(2 C_2(R_{F3})+10\big),\\[1.5ex]
C_{32}&=& \frac{9}{2}+ 2 N_F \,S_2(R_{F3}) \,C_2(R_{F2}) \,d(R_{F2}),\\[1.5ex]
C_{31}&=&\frac{11}{10} +\frac{6}{5} N_F\, S_2(R_{F3}) d(R_{F2}) Y_F^2,\\[1.5ex]
C_{23}&=&12+2 N_F\,S_2(R_{F2})\, C_2(R_{F3})\, d(R_{F3}),  \\[1.5ex]
C_{22}&=&\frac{35}{6}+ N_F\, S_2(R_{F2}) \,d(R_{F3})\big( 2 C_2(R_{F2})+  \frac{20}{3}\big),\\[1.5ex]
C_{21}&=&\frac{9}{10} +\frac{6}{5}  N_F\, S_2(R_{F2}) d(R_{F3}) Y_F^2, \\[1.5ex]
C_{13}&=& \frac{44}{4} +\frac{6}{5} N_F\, C_2(R_{F3}) d(R_{F3})  d(R_{F2})  Y_F^2, \\[1.5ex]
C_{12}&=& \frac{27}{10} +\frac{6}{5} N_F\, C_2(R_{F2}) d(R_{F2}) d(R_{F3}) Y_F^2, \\[1.5ex]
C_{11}&=&\frac{199}{50} + \frac{18}{25} N_F\, d(R_{F3}) d(R_{F2}) Y_F^4.
\eea

\section{Decomposition of the irreducible $SU(5)$ representations}\label{su5}

In this Appendix we collected the branching rules for the embedding $SU(5)\supset SU(3)\times SU(2)\times U(1)$\cite{SLANSKY19811},
\bea\label{su5reps}
{\bf 5}&= & \left({\bf 1}, {\bf 2}, \textstyle\frac{1}{2}\right)\oplus \left({\bf 3}, {\bf 1}, -\textstyle\frac{1}{3}\right),\\
{\bf 10}&= & \left({\bf 1}, {\bf 1}, 1\right)\oplus \left({\bf\bar{3}}, {\bf 1}, -\textstyle\frac{2}{3}\right)\oplus \left({\bf 3}, {\bf 2}, \textstyle\frac{1}{6}\right),\nonumber\\
{\bf 15}&= & \left({\bf 1}, {\bf 3}, 1\right)\oplus \left({\bf 3}, {\bf 2}, \textstyle\frac{1}{6}\right)\oplus \left({\bf 6}, {\bf 1}, -\textstyle\frac{2}{3}\right),\nonumber\\
{\bf 24}&= & \left({\bf 1}, {\bf 1}, 0\right)\oplus \left({\bf 1}, {\bf 3}, 0\right)\oplus \left({\bf 8}, {\bf 1}, 0\right)\oplus\left({\bf 3}, {\bf 2}, -\textstyle\frac{5}{6}\right)\oplus\left({\bf\bar 3} , {\bf 2}, \textstyle\frac{5}{6}\right),\nonumber\\
{\bf 35}&= & \left({\bf 1}, {\bf 4}, -\textstyle\frac{3}{2}\right)\oplus \left({\bf\bar 3}, {\bf 3}, -\textstyle\frac{2}{3}\right)\oplus \left({\bf\bar 6}, {\bf 2}, \textstyle\frac{1}{6}\right)\oplus \left({\bf\bar 10}, {\bf 1}, 1\right),\nonumber\\
{\bf 40}&= & \left({\bf 1}, {\bf 2}, -\textstyle\frac{3}{2}\right)\oplus \left({\bf 3}, {\bf 2}, \textstyle\frac{1}{6}\right)\oplus \left({\bf\bar 3}, {\bf 1}, -\textstyle\frac{2}{3}\right)\oplus \left({\bf\bar 3}, {\bf 3}, -\textstyle\frac{2}{3}\right)\oplus \left({\bf 8}, {\bf 1}, 1\right)\oplus \left({\bf\bar 6}, {\bf 2}, \textstyle\frac{1}{6}\right),\nonumber\\
{\bf 45}&=& \left({\bf 1}, {\bf 2}, \textstyle\frac{1}{2}\right)\oplus \left({\bf 3}, {\bf 1}, -\textstyle\frac{1}{3}\right)\oplus \left({\bf 3}, {\bf 3}, -\textstyle\frac{1}{3}\right)\oplus \left({\bf\bar 3}, {\bf 1}, \textstyle\frac{4}{3}\right)\oplus \left({\bf\bar 3}, {\bf 2}, -\textstyle\frac{7}{6}\right)\oplus \left({\bf\bar 6}, {\bf 1}, -\textstyle\frac{1}{3}\right)\oplus \left({\bf 8}, {\bf 2}, \textstyle\frac{1}{2}\right),\nonumber\\
{\bf 50}&= & \left({\bf 1}, {\bf 1}, -2\right)\oplus \left({\bf 3}, {\bf 1}, -\textstyle\frac{1}{3}\right)\oplus \left({\bf\bar 3}, {\bf 2}, -\textstyle\frac{7}{6}\right)\oplus \left({\bf\bar 6}, {\bf 3}, -\textstyle\frac{1}{3}\right)\oplus \left({\bf 6}, {\bf 1}, \textstyle\frac{4}{3}\right)\oplus \left({\bf 8}, {\bf 2}, \textstyle\frac{1}{2}\right),\nonumber\\
{\bf 70}&= & \left({\bf 1}, {\bf 2}, \textstyle\frac{1}{2}\right)\oplus \left({\bf 1}, {\bf 4}, \textstyle\frac{1}{2}\right)\oplus \left({\bf 3}, {\bf 1}, -\textstyle\frac{1}{3}\right)\oplus \left({\bf 3}, {\bf 3}, -\textstyle\frac{1}{3}\right)\oplus \left({\bf\bar 3}, {\bf 3}, \textstyle\frac{4}{3}\right)\oplus \left({\bf 6}, {\bf 2}, -\textstyle\frac{7}{6}\right)\oplus \left({\bf 8}, {\bf 2}, \textstyle\frac{1}{2}\right)\oplus \left({\bf 15}, {\bf 1}, -\textstyle\frac{1}{3}\right),\nonumber\\
{\bf 70^{\prime}}&= &\left({\bf 1}, {\bf 5}, -2\right)\oplus \left({\bf\bar 3}, {\bf 4}, -\textstyle\frac{7}{6}\right)\oplus \left({\bf\bar 6}, {\bf 3}, -\textstyle\frac{1}{3}\right)\oplus \left({\bf \overline{10}}, {\bf 2}, \textstyle\frac{1}{2}\right)\oplus \left({\bf\overline{15}^{\prime}}, {\bf 1}, \textstyle\frac{4}{3}\right),\nonumber\\
{\bf 75^{\prime}}&= &\left({\bf 1}, {\bf 1}, 0\right)\oplus \left({\bf 3}, {\bf 1}, \textstyle\frac{5}{3}\right)\oplus \left({\bf 3}, {\bf 2}, -\textstyle\frac{5}{6}\right)\oplus \left({\bf\bar 3}, {\bf 1}, -\textstyle\frac{5}{3}\right)\oplus \left({\bf\bar 3}, {\bf 2}, \textstyle\frac{5}{6}\right)\oplus \left({\bf\bar 6}, {\bf 2}, -\textstyle\frac{5}{6}\right)\oplus \left({\bf 6}, {\bf 2}, \textstyle\frac{5}{6}\right)\oplus\nonumber\\
&&\left({\bf 8}, {\bf 1}, 0\right)\oplus \left({\bf 8}, {\bf 3}, 0\right).\nonumber
\eea
\bigskip

\bibliographystyle{utphysmcite}	
\bibliography{myref}


\end{document}

%% file: macros.tex
\newcommand{\newc}{\newcommand*}

\long\def\begincomment#1\endcomment{%
        \begingroup\sf\baselineskip12pt#1\endgroup}

\newc{\etal}{\textrm{et al.}} 
\newc{\eg}{\textrm{e.g.}} 
\newc{\ie}{\textrm{i.e.}}
\newc{\etc}{\textrm{etc}}
\newc\vs{\textrm{vs.}}
\newc{\cl}{\rm {C.L.}}

\newc{\ev}{\ensuremath{\,\mathrm{eV}}}
\newc{\kev}{\ensuremath{\,\mathrm{keV}}}
\newc{\mev}{\ensuremath{\,\mathrm{MeV}}}
\newc{\gev}{\ensuremath{\,\mathrm{GeV}}}
\newc{\tev}{\ensuremath{\,\mathrm{TeV}}}
\newc{\MeV}{\mev} 
\newc{\TeV}{\tev}
\newc{\invpb}{\ensuremath{/\text{pb}}}
\newc{\invfb}{\ensuremath{/\text{fb}}}
\newc\nb{\ensuremath{\,\mathrm{nb}}} \newc\pb{\ensuremath{\,\mathrm{pb}}} \newc\fb{\ensuremath{\,\mathrm{fb}}}
\newc\pc{\ensuremath{\,\mathrm{pc}}}
\newc\kpc{\ensuremath{\,\mathrm{kpc}}}
\newc\mpc{\ensuremath{\,\mathrm{Mpc}}}
\newc\ps{\ensuremath{\,\mathrm{ps}}} 
\newc\cmeter{\ensuremath{\,\mathrm{cm}}} 
\newc\meter{\ensuremath{\,\mathrm{m}}} 
\newc\kmeter{\ensuremath{\,\mathrm{km}}}
\newc\second{\ensuremath{\,\mathrm{s}}}
\newc\msecond{\ensuremath{\,\mathrm{ms}}}
\newc\nsecond{\ensuremath{\,\mathrm{ns}}}
\newc\psecond{\ensuremath{\,\mathrm{ps}}}

\newc{\chisqmin}{\ensuremath{\chi^2_{\mathrm{min}}}}
\newc{\Delchisq}{\ensuremath{\Delta\chi^2}}
\newc{\chisq}{\ensuremath{\chi^2}}
\newc{\like}{\ensuremath{\mathcal{L}}}

\newc\lsim{\ensuremath{\mathrel{\rlap{\lower4pt\hbox{\hskip1pt$\sim$}}\raise1pt\hbox{$<$}}}}
\newc\gsim{\ensuremath{\mathrel{\rlap{\lower4pt\hbox{\hskip1pt$\sim$}}\raise1pt\hbox{$>$}}}}
\newc{\VEV}[1]{\ensuremath{\langle #1 \rangle}}
\newc{\dl}{\ensuremath{\stackrel{\leftarrow}{D}}}
\newc{\dr}{\ensuremath{\stackrel{\rightarrow}{D}}}

\newc{\bcenter}{\begin{center}}    \newc{\ecenter}{\end{center}}
\newc{\bfl}{\begin{flushleft}}    \newc{\efl}{\end{flushleft}}
\newc{\bfr}{\begin{flushright}}    \newc{\efr}{\end{flushright}}

\newc{\bi}{\begin{itemize}}
\newc{\ei}{\end{itemize}}
\newc{\bed}{\begin{description}}
\newc{\eed}{\end{description}}
\newc{\ben}{\begin{enumerate}}
\newc{\een}{\end{enumerate}}

\newc{\be}{\begin{equation}}
\newc{\ee}{\end{equation}}
\newc{\bea}{\begin{eqnarray}}
\newc{\eea}{\end{eqnarray}}
\newc{\bfle}{\begin{flalign}}
\newc{\efle}{\end{flalign}}
\newc{\ra}{\rightarrow}

\newc{\alphas}{\ensuremath{\alpha_s}}
\newc{\alphatwo}{\ensuremath{\alpha_2}}
\newc{\alphaone}{\ensuremath{\alpha_1}}
\newc{\alphai}[1]{\ensuremath{\alpha_{#1}}}
\newc{\alphaem}{\ensuremath{\alpha_{\mathrm{em}}}}
\newc{\alphaeff}{\ensuremath{\alpha_{\mathrm{eff}}}}
\newc{\sineff}{\ensuremath{\sin \theta_{\mathrm{eff}}}}
\newc{\sinsqeff}{\ensuremath{\sin^2 \theta_{\mathrm{eff}}}}
\newc{\dalphahad}{\ensuremath{\Delta \alpha_{\mathrm{had}}}}
\newc{\yt}{\ensuremath{h_t}} \newc{\yb}{\ensuremath{h_b}} \newc{\ytau}{\ensuremath{h_{\tau}}}
\newc\mz{\ensuremath{M_Z}} 
\newc\mw{\ensuremath{m_W}}
\newc\mZ{\mz}        \newc\mW{\mw}
\newc\mhsm{\ensuremath{ m_{H_{\mathrm{SM}}}}}
\newc{\mtop}{\ensuremath{ m_t}}               \newc{\mtpole}{\ensuremath{ M_t}}
\newc{\mbottom}{\ensuremath{ m_b}} 
\newc{\mtau}{\ensuremath{ m_{\tau}}}
\newc{\mt}{\mtpole}
\newc{\mb}{\mbottom} 
\newc{\rtwogg}{\ensuremath{R_{h_2}(\gamma\gamma)}}
\newc{\rtwozz}{\ensuremath{R_{h_2}(ZZ)}}
\newc{\ronegg}{\ensuremath{R_{h_1}(\gamma\gamma)}}
\newc{\ronezz}{\ensuremath{R_{h_1}(ZZ)}}
\newc{\rsiggg}{\ensuremath{R_{h_\textrm{sig}}(\gamma\gamma)}}
\newc{\rsigzz}{\ensuremath{R_{h_\textrm{sig}}(ZZ)}}
\newc{\llbar}{\ensuremath{\ell\bar{\ell}}}
\newc{\tauptaum}{\ensuremath{ \tau^+\tau^-}}
\newc{\qqbar}{\ensuremath{ q\bar{q}}} \newc{\ppbar}{\ensuremath{ p\bar{p}}}
\newc{\bbbar}{\ensuremath{ b\bar{b}}} \newc{\ttbar}{\ensuremath{ t\bar{t}}}
\newc{\ffbar}{\ensuremath{ f\bar{f}}} \newc{\tautaubar}{\ensuremath{ \tau\bar{\tau}}}

\newc{\mchi}{\ensuremath{m_\neutone}}
\newc{\squark}{\ensuremath{\tilde{q}}}
\newc{\slepton}{\ensuremath{\tilde{l}}}
\newc{\gluino}{\ensuremath{\tilde{g}}} 
\newc{\mgluino}{\ensuremath{{m_{\gluino}}}}
\newc{\wino}{\ensuremath{\tilde{W}}} 
\newc{\mwino}{\ensuremath{{m_{\wino}}}}
\newc{\tone}{\ensuremath{{\tilde{t}_1}}}
\newc{\Hone}{\ensuremath{{\tilde{H}_{1}}}}
\newc{\Htwo}{\ensuremath{{\tilde{H}_{2}}}}
\newc{\Hhtwo}{\ensuremath{{H_{2}}}}
\newc{\qli}{\ensuremath{{\tilde{Q}_{i}}}}
\newc{\uri}{\ensuremath{{\tilde{u}_{i}}}}
\newc{\dri}{\ensuremath{{\tilde{d}_{i}}}}
\newc{\lli}{\ensuremath{{\tilde{L}_{i}}}}
\newc{\eri}{\ensuremath{{\tilde{e}_{i}}}}

\newc{\sthw}{\ensuremath{ \sin\theta_W}}              \newc{\cthw}{\ensuremath{\cos\theta_W}}
\newc{\tanthw}{\ensuremath{ \tan\theta_W}}              \newc{\cotthw}{\ensuremath{\cot\theta_W}}
\newc{\ssqthw}{\ensuremath{\sin^2 \theta_W}}
\newc{\msbar}{\ensuremath{\overline{MS}}} \newc{\drbar}{\ensuremath{\overline{DR}}}
\newc{\mtmtsmmsbar}{\ensuremath{ m_t(m_t)^{\msbar}_{{\mathrm{SM}}}}}
\newc{\mtmtsmdrbar}{\ensuremath{ m_t(m_t)^{\drbar}_{{\mathrm{SM}}}}}
\newc{\mtmtmssmdrbar}{\ensuremath{ m_t(m_t)^{\drbar}_{{\mathrm{SUSY}}}}}
\newc{\mbmbmsbar}{\ensuremath{ m_b(m_b)^{\msbar} }}
\newc{\mbmbsmmsbar}{\ensuremath{ m_b(m_b)^{\msbar}_{{\mathrm{SM}}}}}
\newc{\mbmzsmmsbar}{\ensuremath{ m_b(\mz)^{\msbar}_{{\mathrm{SM}}}}}
\newc{\mbmzsmdrbar}{\ensuremath{ m_b(\mz)^{\drbar}_{{\mathrm{SM}}}}}
\newc{\mbmzmssmdrbar}{\ensuremath{ m_b(\mz)^{\drbar}_{{\mathrm{SUSY}}}}}
\newc{\mtaumzsmmsbar}{\ensuremath{ m_{\tau}(\mz)^{\msbar}_{{\mathrm{SM}}}}}
\newc{\mtaumzsmdrbar}{\ensuremath{ m_{\tau}(\mz)^{\drbar}_{{\mathrm{SM}}}}}
\newc{\mtaumzmssmdrbar}{\ensuremath{ m_{\tau}(\mz)^{\drbar}_{{\mathrm{SUSY}}}}}
\newc{\alphasmzms}{\ensuremath{\alpha_s(M_Z)^{\overline{MS}}}}
\newc{\alphaimzms}[1]{\ensuremath{\alpha_{#1}(M_Z)^{\overline{MS}}}}
\newc{\alphaemmz}{\ensuremath{\alpha_{\mathrm{em}}(M_Z)^{\overline{MS}}}}

\newc{\mzero}{\ensuremath{{m_0}}}
\newc{\mhalf}{\ensuremath{ m_{1/2}}}
\newc{\tanb}{\ensuremath{\tan\beta}}
\newc{\azero}{\ensuremath{ A_0}}
\newc{\signmu}{\ensuremath{\rm{sgn}\,\mu}}
\newc{\atau}{\ensuremath{{A_{\tau}}}}
\newc{\mueff}{\ensuremath{\mu_{\rm{eff}}}}
\newc{\lam}{\ensuremath{{\lambda}}}
\newc{\kap}{\ensuremath{{\kappa}}}
\newc{\alam}{\ensuremath{{A_{\lambda}}}}
\newc{\akap}{\ensuremath{{A_{\kappa}}}}
\newc{\hs}{\ensuremath{ H_s}}      
\newc{\mhs}{\ensuremath{ m_{H_s}}} 
\newc{\mgut}{\ensuremath{ M_{\rm GUT}}}
\newc{\mvl}{\ensuremath{ M_{\rm VL}}}
\newc{\gut}{\ensuremath{{\rm GUT}}}
\newc{\mplanck}{\ensuremath{ M_{\rm P}}}      \newc{\mpl}{\ensuremath{ M_{\rm Pl}}}
\newc{\msusy}{\ensuremath{ M_{\rm SUSY}}}      \newc{\ms}{\ensuremath{ M_{\rm S}}}
 \newc{\hu}{\ensuremath{ H_u}}       \newc{\hd}{\ensuremath{ H_d}}
 \newc{\mhu}{\ensuremath{ m_{H_u}}}       \newc{\mhd}{\ensuremath{ m_{H_d}}}
 \newc{\mhuew}{\ensuremath{ m^{\ast}_{H_u}}}       \newc{\mhdew}{\ensuremath{ m^{\ast}_{H_d}}}
 \newc{\mhuewsq}{\ensuremath{ m^{\ast\, 2}_{H_u}}}       \newc{\mhdewsq}{\ensuremath{ m^{\ast\, 2}_{H_d}}}
 \newc{\mhl}{\ensuremath{m_\hl}} 
 \newc{\mhone}{\ensuremath{m_{h_1}}} 
 \newc{\mhtwo}{\ensuremath{m_{h_2}}} 
 \newc{\mhi}{\ensuremath{m_{\tilde{h}}}} 
 \newc{\mul}{\ensuremath{m_{\tilde{u}_L}}} 
 \newc{\mtone}{\ensuremath{m_{\tilde{t}_1}}} 
 \newc{\ma}{\ensuremath{m_A}} 
 \newc{\mH}{\ensuremath{m_H}} 
 \newc{\maone}{\ensuremath{m_{a_1}}} 
 \newc{\matwo}{\ensuremath{m_{a_2}}}
 \newc{\hone}{\ensuremath{h_1}}
 \newc{\htwo}{\ensuremath{h_2}}
 \newc{\aone}{\ensuremath{a_1}}
 \newc{\atwo}{\ensuremath{a_2}}
 \newc{\mqthree}{\ensuremath{m_{\tilde{Q}_3}^2}}
 \newc{\muthree}{\ensuremath{m_{\tilde{u}_3}^2}}
 \newc{\mqli}{\ensuremath{m_{\tilde{Q}_{i}}}}
 \newc{\muri}{\ensuremath{m_{\tilde{u}_{i}}}}
 \newc{\mdri}{\ensuremath{m_{\tilde{d}_{i}}}}
 \newc{\mlli}{\ensuremath{m_{\tilde{L}_{i}}}}
 \newc{\meri}{\ensuremath{m_{\tilde{e}_{i}}}}
 \newc{\ts}{\ensuremath{T_{SUSY}}}

\newc{\sigsip}{\ensuremath{\sigma^{\rm SI}_{p}}}	\newc{\sigsin}{\ensuremath{\sigma^{\rm SI}_{n}}}
\newc{\sigsdp}{\ensuremath{\sigma^{\rm SD}_{p}}}	\newc{\sigsdn}{\ensuremath{\sigma^{\rm SD}_{n}}}
\newc{\sigsi}{\ensuremath{\sigma^{\rm SI}}}	\newc{\sigsd}{\ensuremath{\sigma^{\rm SD}}}
\newc{\abund}{\ensuremath{ \Omega h^2}}
\newc{\omegadm}{\ensuremath{ \Omega_{{\rm DM}}}}     \newc{\abunddm}{\ensuremath{ \Omega_{{\rm DM}} h^2}} 
\newc{\omegam}{\ensuremath{ \Omega_{{\rm m}}}}       \newc{\abundm}{\ensuremath{ \Omega_{{\rm m}} h^2}}
\newc{\omegab}{\ensuremath{ \Omega_{{\rm b}}}}	\newc{\abundb}{\ensuremath{ \Omega_{{\rm b}} h^2}}
\newc{\omegatot}{\ensuremath{ \Omega_{{\rm TOT}}}}
\newc{\omegacdm}{\ensuremath{ \Omega_{{\rm CDM}}}}   \newc{\abundcdm}{\ensuremath{ \Omega_{{\rm CDM}} h^2}}
\newc{\omegalambda}{\ensuremath{ \Omega_{\Lambda}}} \newc{\abundlambda}{\ensuremath{ \Omega_{\Lambda} h^2}}
\newc{\omegarad}{\ensuremath{ \Omega_{{\rm rad}}}}  \newc{\abundrad}{\ensuremath{ \Omega_{{\rm rad}} h^2}}
\newc{\rhocrit}{\ensuremath{ \rho_{\rm crit}}}
\newc{\rhochi}{\ensuremath{ \rho_{\chi}}}
\newc{\abunchi}{\ensuremath{\Omega_\chi h^2}}
\newc{\abundlsp}{\ensuremath{\Omega_{\rm LSP}h^2}}

\newc{\amu}{\ensuremath{ a_{\mu}}}        \newc{\amususy}{\ensuremath{ a_{\mu}^{\mathrm{SUSY}}}}
\newc{\amuexpt}{\ensuremath{ a_{\mu}^{\mathrm{expt}}}}        \newc{\amusm}{\ensuremath{ a_{\mu}^{\mathrm{SM}}}}
\newc\deltaamu{\ensuremath{\Delta a_{\mu}}} \newc{\deltaamususy}{\ensuremath{\delta a_{\mu}^{\mathrm{SUSY}}}}
\newc\gmtwo{\ensuremath{ (g-2)_{\mu}}} 
\newc{\deltagmtwomususy}{\ensuremath{\delta\left(g-2\right)_{\mu}^{\mathrm{SUSY}}}}
\newc{\deltagmtwomu}{\ensuremath{\delta\left(g-2\right)_{\mu}}}
\newc\BR{\ensuremath{\rm BR}}
\newc\bsgamma{\ensuremath{ b\rightarrow s \gamma }}
\newc\bxsgamma{\ensuremath{\overline{B}\rightarrow X_{s}\gamma}}
\newc\brbsgamma{\ensuremath{\BR\left(\bsgamma\right)}}
\newc\brbxsgamma{\ensuremath{\BR\left(\bxsgamma\right)}}
\newc\bsmumu{\ensuremath{B_s\to\mu^+\mu^-}}
\newc\brbsmumu{\ensuremath{\BR\left(B_s\to\mu^+\mu^-\right)}}
\newc\bdmmumu{\ensuremath{\overline{B}_d\to\mu^+\mu^-}}
\newc\bbbarmix{\ensuremath{\overline{B}_s\mbox{-}B_s}}      
\newc\delmbs{\ensuremath{\Delta M_{B_s}}}
\newc{\butaunu}{\ensuremath{B_u \rightarrow \tau \nu}}
\newc{\brbutaunu}{\ensuremath{\BR\left(B_u \rightarrow \tau \nu\right)}}

\newcommand*{\reftable}[1]{Table~\ref{#1}}         \newcommand*{\reftables}[2]{Tables~\ref{#1} and \ref{#2}}
       
\newcommand*{\reffig}[1]{Fig.~\ref{#1}}
\newcommand*{\reffigs}[2]{Figs.~\ref{#1} and \ref{#2}} 
        \newcommand*{\refeq}[1]{Eq.~\ref{#1}}
     \newcommand*{\refsec}[1]{Sec.~\ref{#1}}
            
\newcommand*{\refap}[1]{Appendix~\ref{#1}}     
\newcommand*{\refeqst}[2]{Eq.~\ref{#1}-\ref{#2}}

\newcommand*{\neutone}{\ensuremath{\tilde{\chi}^0_1}}

\newcommand*{\eight}{\ensuremath{\sqrt{s}=8\tev}}

\newcommand*{\madgr}{\texttt{MadGraph5\_aMC@NLO}}

\let\oldcite\cite
\renewcommand*{\cite}{~\oldcite}

\newcommand*{\hl}{\ensuremath{h}}


%% file: arXiv_v2.bbl
\ifx\mcitethebibliography\mciteundefinedmacro
\PackageError{unsrtM.bst}{mciteplus.sty has not been loaded}
{This bibstyle requires the use of the mciteplus package.}\fi
\begin{mcitethebibliography}{10}

\bibitem{Georgi:1974sy}
H.~Georgi and S.~L. Glashow, ``{Unity of All Elementary Particle Forces},''
\href{http://dx.doi.org/10.1103/PhysRevLett.32.438}{{\em Phys. Rev. Lett.}
  {\bfseries 32} (1974) 438--441}.
\mciteBstWouldAddEndPunctfalse
\mciteSetBstMidEndSepPunct{\mcitedefaultmidpunct}
{}{\mcitedefaultseppunct}\relax
\EndOfBibitem
\bibitem{Pati:1974yy}
J.~C. Pati and A.~Salam, ``{Lepton Number as the Fourth Color},''
  \href{http://dx.doi.org/10.1103/PhysRevD.10.275,
  10.1103/PhysRevD.11.703.2}{{\em Phys. Rev.} {\bfseries D10} (1974) 275--289}.
[Erratum: Phys. Rev.D11,703(1975)].
\mciteBstWouldAddEndPunctfalse
\mciteSetBstMidEndSepPunct{\mcitedefaultmidpunct}
{}{\mcitedefaultseppunct}\relax
\EndOfBibitem
\bibitem{Mohapatra:1974hk}
R.~N. Mohapatra and J.~C. Pati, ``{Left-Right Gauge Symmetry and an
  Isoconjugate Model of CP Violation},''
\href{http://dx.doi.org/10.1103/PhysRevD.11.566}{{\em Phys. Rev.} {\bfseries
  D11} (1975) 566--571}.
\mciteBstWouldAddEndPunctfalse
\mciteSetBstMidEndSepPunct{\mcitedefaultmidpunct}
{}{\mcitedefaultseppunct}\relax
\EndOfBibitem
\bibitem{Fritzsch:1974nn}
H.~Fritzsch and P.~Minkowski, ``{Unified Interactions of Leptons and
  Hadrons},''
\href{http://dx.doi.org/10.1016/0003-4916(75)90211-0}{{\em Annals Phys.}
  {\bfseries 93} (1975) 193--266}.
\mciteBstWouldAddEndPunctfalse
\mciteSetBstMidEndSepPunct{\mcitedefaultmidpunct}
{}{\mcitedefaultseppunct}\relax
\EndOfBibitem
\bibitem{Georgi:1974my}
H.~Georgi, ``{The State of the Art—Gauge Theories},''
\href{http://dx.doi.org/10.1063/1.2947450}{{\em AIP Conf. Proc.} {\bfseries 23}
  (1975) 575--582}.
\mciteBstWouldAddEndPunctfalse
\mciteSetBstMidEndSepPunct{\mcitedefaultmidpunct}
{}{\mcitedefaultseppunct}\relax
\EndOfBibitem
\bibitem{Cook:1979pr}
G.~P. Cook, K.~T. Mahanthappa, and M.~A. Sher, ``{Effects of Heavy Colored
  Higgs Scalars on SU(5) Predictions},''
\href{http://dx.doi.org/10.1016/0370-2693(80)90957-0}{{\em Phys. Lett.}
  {\bfseries 90B} (1980) 398--400}.
\mciteBstWouldAddEndPunctfalse
\mciteSetBstMidEndSepPunct{\mcitedefaultmidpunct}
{}{\mcitedefaultseppunct}\relax
\EndOfBibitem
\bibitem{Dixit:1989ff}
V.~V. Dixit and M.~Sher, ``{The Futility of High Precision SO(10)
  Calculations},''
\href{http://dx.doi.org/10.1103/PhysRevD.40.3765}{{\em Phys. Rev.} {\bfseries
  D40} (1989) 3765}.
\mciteBstWouldAddEndPunctfalse
\mciteSetBstMidEndSepPunct{\mcitedefaultmidpunct}
{}{\mcitedefaultseppunct}\relax
\EndOfBibitem
\bibitem{Langacker:1992rq}
P.~Langacker and N.~Polonsky, ``{Uncertainties in coupling constant
  unification},'' \href{http://dx.doi.org/10.1103/PhysRevD.47.4028}{{\em Phys.
  Rev.} {\bfseries D47} (1993) 4028--4045},
\href{http://arxiv.org/abs/hep-ph/9210235}{{\ttfamily arXiv:hep-ph/9210235
  [hep-ph]}}.
\mciteBstWouldAddEndPunctfalse
\mciteSetBstMidEndSepPunct{\mcitedefaultmidpunct}
{}{\mcitedefaultseppunct}\relax
\EndOfBibitem
\bibitem{Hagiwara:1992ys}
K.~Hagiwara and Y.~Yamada, ``{Grand unification threshold effects in
  supersymmetric SU(5) models},''
\href{http://dx.doi.org/10.1103/PhysRevLett.70.709}{{\em Phys. Rev. Lett.}
  {\bfseries 70} (1993) 709--712}.
\mciteBstWouldAddEndPunctfalse
\mciteSetBstMidEndSepPunct{\mcitedefaultmidpunct}
{}{\mcitedefaultseppunct}\relax
\EndOfBibitem
\bibitem{Schwichtenberg:2018cka}
J.~Schwichtenberg, ``{Gauge Coupling Unification without Supersymmetry},''
  \href{http://dx.doi.org/10.1140/epjc/s10052-019-6878-1}{{\em Eur. Phys. J.}
  {\bfseries C79} no.~4, (2019) 351},
\href{http://arxiv.org/abs/1808.10329}{{\ttfamily arXiv:1808.10329 [hep-ph]}}.
\mciteBstWouldAddEndPunctfalse
\mciteSetBstMidEndSepPunct{\mcitedefaultmidpunct}
{}{\mcitedefaultseppunct}\relax
\EndOfBibitem
\bibitem{Achiman:1979}
Y.~Achiman and B.~Stech, {\em New Phenomena in Lepton-Hadron Physics}.
\newblock edited by D.E.C. Fries and J. Wess, Plenum, New York, 1979, p.
  303.\relax
\mciteBstWouldAddEndPunctfalse
\mciteSetBstMidEndSepPunct{\mcitedefaultmidpunct}
{}{\mcitedefaultseppunct}\relax
\EndOfBibitem
\bibitem{Rujula:1984}
A.~de~R\'ujula, H.~Georgi, and S.~L. Glashow, {\em Fifth Workshop on Grand
  Unification}.
\newblock edited by K. Kang, H. Fried, and P. Frampton, World Scientific,
  Singapore, 1984, p. 88.\relax
\mciteBstWouldAddEndPunctfalse
\mciteSetBstMidEndSepPunct{\mcitedefaultmidpunct}
{}{\mcitedefaultseppunct}\relax
\EndOfBibitem
\bibitem{Perez:2014kfa}
P.~Fileviez~Perez and S.~Ohmer, ``{Low Scale Unification of Gauge
  Interactions},'' \href{http://dx.doi.org/10.1103/PhysRevD.90.037701}{{\em
  Phys. Rev.} {\bfseries D90} no.~3, (2014) 037701},
\href{http://arxiv.org/abs/1405.1199}{{\ttfamily arXiv:1405.1199 [hep-ph]}}.
\mciteBstWouldAddEndPunctfalse
\mciteSetBstMidEndSepPunct{\mcitedefaultmidpunct}
{}{\mcitedefaultseppunct}\relax
\EndOfBibitem
\bibitem{Rizzo:1991tc}
T.~G. Rizzo, ``{Desert guts and new light degrees of freedom},''
\href{http://dx.doi.org/10.1103/PhysRevD.45.R3903}{{\em Phys. Rev.} {\bfseries
  D45} (1992) 3903--3905}.
\mciteBstWouldAddEndPunctfalse
\mciteSetBstMidEndSepPunct{\mcitedefaultmidpunct}
{}{\mcitedefaultseppunct}\relax
\EndOfBibitem
\bibitem{Zhang:2000zy}
B.~Zhang and H.-Q. Zheng, ``{Top quark, heavy fermions and the composite Higgs
  boson},'' \href{http://dx.doi.org/10.1088/0253-6102/35/2/162}{{\em Commun.
  Theor. Phys.} {\bfseries 35} (2001) 162--166},
\href{http://arxiv.org/abs/hep-ph/0003065}{{\ttfamily arXiv:hep-ph/0003065
  [hep-ph]}}.
\mciteBstWouldAddEndPunctfalse
\mciteSetBstMidEndSepPunct{\mcitedefaultmidpunct}
{}{\mcitedefaultseppunct}\relax
\EndOfBibitem
\bibitem{Choudhury:2001hs}
D.~Choudhury, T.~M.~P. Tait, and C.~E.~M. Wagner, ``{Beautiful mirrors and
  precision electroweak data},''
  \href{http://dx.doi.org/10.1103/PhysRevD.65.053002}{{\em Phys. Rev.}
  {\bfseries D65} (2002) 053002},
\href{http://arxiv.org/abs/hep-ph/0109097}{{\ttfamily arXiv:hep-ph/0109097
  [hep-ph]}}.
\mciteBstWouldAddEndPunctfalse
\mciteSetBstMidEndSepPunct{\mcitedefaultmidpunct}
{}{\mcitedefaultseppunct}\relax
\EndOfBibitem
\bibitem{Li:2003zh}
L.-F. Li and F.~Wu, ``{Coupling constant unification in extensions of standard
  model},'' \href{http://dx.doi.org/10.1142/S0217751X04019391}{{\em Int. J.
  Mod. Phys.} {\bfseries A19} (2004) 3217--3224},
\href{http://arxiv.org/abs/hep-ph/0304238}{{\ttfamily arXiv:hep-ph/0304238
  [hep-ph]}}.
\mciteBstWouldAddEndPunctfalse
\mciteSetBstMidEndSepPunct{\mcitedefaultmidpunct}
{}{\mcitedefaultseppunct}\relax
\EndOfBibitem
\bibitem{Morrissey:2003sc}
D.~E. Morrissey and C.~E.~M. Wagner, ``{Beautiful mirrors, unification of
  couplings and collider phenomenology},''
  \href{http://dx.doi.org/10.1103/PhysRevD.69.053001}{{\em Phys. Rev.}
  {\bfseries D69} (2004) 053001},
\href{http://arxiv.org/abs/hep-ph/0308001}{{\ttfamily arXiv:hep-ph/0308001
  [hep-ph]}}.
\mciteBstWouldAddEndPunctfalse
\mciteSetBstMidEndSepPunct{\mcitedefaultmidpunct}
{}{\mcitedefaultseppunct}\relax
\EndOfBibitem
\bibitem{Giudice:2004tc}
G.~F. Giudice and A.~Romanino, ``{Split supersymmetry},''
  \href{http://dx.doi.org/10.1016/j.nuclphysb.2004.11.048,
  10.1016/j.nuclphysb.2004.08.001}{{\em Nucl. Phys.} {\bfseries B699} (2004)
  65--89}, \href{http://arxiv.org/abs/hep-ph/0406088}{{\ttfamily
  arXiv:hep-ph/0406088 [hep-ph]}}.
[Erratum: Nucl. Phys.B706,487(2005)].
\mciteBstWouldAddEndPunctfalse
\mciteSetBstMidEndSepPunct{\mcitedefaultmidpunct}
{}{\mcitedefaultseppunct}\relax
\EndOfBibitem
\bibitem{Dorsner:2005fq}
I.~Dorsner and P.~Fileviez~Perez, ``{Unification without supersymmetry:
  Neutrino mass, proton decay and light leptoquarks},''
  \href{http://dx.doi.org/10.1016/j.nuclphysb.2005.06.016}{{\em Nucl. Phys.}
  {\bfseries B723} (2005) 53--76},
\href{http://arxiv.org/abs/hep-ph/0504276}{{\ttfamily arXiv:hep-ph/0504276
  [hep-ph]}}.
\mciteBstWouldAddEndPunctfalse
\mciteSetBstMidEndSepPunct{\mcitedefaultmidpunct}
{}{\mcitedefaultseppunct}\relax
\EndOfBibitem
\bibitem{EmmanuelCosta:2005nh}
D.~Emmanuel-Costa and R.~Gonzalez~Felipe, ``{Minimal string-scale unification
  of gauge couplings},''
  \href{http://dx.doi.org/10.1016/j.physletb.2005.07.038}{{\em Phys. Lett.}
  {\bfseries B623} (2005) 111--118},
\href{http://arxiv.org/abs/hep-ph/0505257}{{\ttfamily arXiv:hep-ph/0505257
  [hep-ph]}}.
\mciteBstWouldAddEndPunctfalse
\mciteSetBstMidEndSepPunct{\mcitedefaultmidpunct}
{}{\mcitedefaultseppunct}\relax
\EndOfBibitem
\bibitem{Shrock:2008sb}
R.~Shrock, ``{Variants of the Standard Model with Electroweak-Singlet
  Quarks},'' \href{http://dx.doi.org/10.1103/PhysRevD.78.076009}{{\em Phys.
  Rev.} {\bfseries D78} (2008) 076009},
\href{http://arxiv.org/abs/0809.0087}{{\ttfamily arXiv:0809.0087 [hep-ph]}}.
\mciteBstWouldAddEndPunctfalse
\mciteSetBstMidEndSepPunct{\mcitedefaultmidpunct}
{}{\mcitedefaultseppunct}\relax
\EndOfBibitem
\bibitem{Gogoladze:2010in}
I.~Gogoladze, B.~He, and Q.~Shafi, ``{New Fermions at the LHC and Mass of the
  Higgs Boson},'' \href{http://dx.doi.org/10.1016/j.physletb.2010.05.076}{{\em
  Phys. Lett.} {\bfseries B690} (2010) 495--500},
\href{http://arxiv.org/abs/1004.4217}{{\ttfamily arXiv:1004.4217 [hep-ph]}}.
\mciteBstWouldAddEndPunctfalse
\mciteSetBstMidEndSepPunct{\mcitedefaultmidpunct}
{}{\mcitedefaultseppunct}\relax
\EndOfBibitem
\bibitem{Dermisek:2012as}
R.~Dermisek, ``{Insensitive Unification of Gauge Couplings},''
  \href{http://dx.doi.org/10.1016/j.physletb.2012.06.037}{{\em Phys. Lett.}
  {\bfseries B713} (2012) 469--472},
\href{http://arxiv.org/abs/1204.6533}{{\ttfamily arXiv:1204.6533 [hep-ph]}}.
\mciteBstWouldAddEndPunctfalse
\mciteSetBstMidEndSepPunct{\mcitedefaultmidpunct}
{}{\mcitedefaultseppunct}\relax
\EndOfBibitem
\bibitem{Dermisek:2012ke}
R.~Dermisek, ``{Unification of gauge couplings in the standard model with extra
  vectorlike families},''
  \href{http://dx.doi.org/10.1103/PhysRevD.87.055008}{{\em Phys. Rev.}
  {\bfseries D87} no.~5, (2013) 055008},
\href{http://arxiv.org/abs/1212.3035}{{\ttfamily arXiv:1212.3035 [hep-ph]}}.
\mciteBstWouldAddEndPunctfalse
\mciteSetBstMidEndSepPunct{\mcitedefaultmidpunct}
{}{\mcitedefaultseppunct}\relax
\EndOfBibitem
\bibitem{Dorsner:2014wva}
I.~Dorsner, S.~Fajfer, and I.~Mustac, ``{Light vector-like fermions in a
  minimal SU(5) setup},''
  \href{http://dx.doi.org/10.1103/PhysRevD.89.115004}{{\em Phys. Rev.}
  {\bfseries D89} no.~11, (2014) 115004},
\href{http://arxiv.org/abs/1401.6870}{{\ttfamily arXiv:1401.6870 [hep-ph]}}.
\mciteBstWouldAddEndPunctfalse
\mciteSetBstMidEndSepPunct{\mcitedefaultmidpunct}
{}{\mcitedefaultseppunct}\relax
\EndOfBibitem
\bibitem{Xiao:2014kba}
M.-L. Xiao and J.-H. Yu, ``{Stabilizing electroweak vacuum in a vectorlike
  fermion model},'' \href{http://dx.doi.org/10.1103/PhysRevD.90.014007,
  10.1103/PhysRevD.90.019901}{{\em Phys. Rev.} {\bfseries D90} no.~1, (2014)
  014007}, \href{http://arxiv.org/abs/1404.0681}{{\ttfamily arXiv:1404.0681
  [hep-ph]}}.
[Addendum: Phys. Rev.D90,no.1,019901(2014)].
\mciteBstWouldAddEndPunctfalse
\mciteSetBstMidEndSepPunct{\mcitedefaultmidpunct}
{}{\mcitedefaultseppunct}\relax
\EndOfBibitem
\bibitem{Bhattacherjee:2017cxh}
B.~Bhattacherjee, P.~Byakti, A.~Kushwaha, and S.~K. Vempati, ``{Unification
  with Vector-like fermions and signals at LHC},''
  \href{http://dx.doi.org/10.1007/JHEP05(2018)090}{{\em JHEP} {\bfseries 05}
  (2018) 090},
\href{http://arxiv.org/abs/1702.06417}{{\ttfamily arXiv:1702.06417 [hep-ph]}}.
\mciteBstWouldAddEndPunctfalse
\mciteSetBstMidEndSepPunct{\mcitedefaultmidpunct}
{}{\mcitedefaultseppunct}\relax
\EndOfBibitem
\bibitem{ArkaniHamed:2004fb}
N.~Arkani-Hamed and S.~Dimopoulos, ``{Supersymmetric unification without low
  energy supersymmetry and signatures for fine-tuning at the LHC},''
  \href{http://dx.doi.org/10.1088/1126-6708/2005/06/073}{{\em JHEP} {\bfseries
  06} (2005) 073},
\href{http://arxiv.org/abs/hep-th/0405159}{{\ttfamily arXiv:hep-th/0405159
  [hep-th]}}.
\mciteBstWouldAddEndPunctfalse
\mciteSetBstMidEndSepPunct{\mcitedefaultmidpunct}
{}{\mcitedefaultseppunct}\relax
\EndOfBibitem
\bibitem{Barger:2006fm}
V.~Barger, J.~Jiang, P.~Langacker, and T.~Li, ``{String scale gauge coupling
  unification with vector-like exotics and non-canonical U(1)(Y)
  normalization},'' \href{http://dx.doi.org/10.1142/S0217751X07038128}{{\em
  Int. J. Mod. Phys.} {\bfseries A22} (2007) 6203--6218},
\href{http://arxiv.org/abs/hep-ph/0612206}{{\ttfamily arXiv:hep-ph/0612206
  [hep-ph]}}.
\mciteBstWouldAddEndPunctfalse
\mciteSetBstMidEndSepPunct{\mcitedefaultmidpunct}
{}{\mcitedefaultseppunct}\relax
\EndOfBibitem
\bibitem{Barger:2007qb}
V.~Barger, N.~G. Deshpande, J.~Jiang, P.~Langacker, and T.~Li, ``{Implications
  of Canonical Gauge Coupling Unification in High-Scale Supersymmetry
  Breaking},'' \href{http://dx.doi.org/10.1016/j.nuclphysb.2007.10.013}{{\em
  Nucl. Phys.} {\bfseries B793} (2008) 307--325},
\href{http://arxiv.org/abs/hep-ph/0701136}{{\ttfamily arXiv:hep-ph/0701136
  [hep-ph]}}.
\mciteBstWouldAddEndPunctfalse
\mciteSetBstMidEndSepPunct{\mcitedefaultmidpunct}
{}{\mcitedefaultseppunct}\relax
\EndOfBibitem
\bibitem{Calibbi:2009cp}
L.~Calibbi, L.~Ferretti, A.~Romanino, and R.~Ziegler, ``{Gauge coupling
  unification, the GUT scale, and magic fields},''
  \href{http://dx.doi.org/10.1016/j.physletb.2009.01.012}{{\em Phys. Lett.}
  {\bfseries B672} (2009) 152--157},
\href{http://arxiv.org/abs/0812.0342}{{\ttfamily arXiv:0812.0342 [hep-ph]}}.
\mciteBstWouldAddEndPunctfalse
\mciteSetBstMidEndSepPunct{\mcitedefaultmidpunct}
{}{\mcitedefaultseppunct}\relax
\EndOfBibitem
\bibitem{Donkin:2010ta}
I.~Donkin and A.~Hebecker, ``{Precision Gauge Unification from Extra Yukawa
  Couplings},'' \href{http://dx.doi.org/10.1007/JHEP09(2010)044}{{\em JHEP}
  {\bfseries 09} (2010) 044},
\href{http://arxiv.org/abs/1007.3990}{{\ttfamily arXiv:1007.3990 [hep-ph]}}.
\mciteBstWouldAddEndPunctfalse
\mciteSetBstMidEndSepPunct{\mcitedefaultmidpunct}
{}{\mcitedefaultseppunct}\relax
\EndOfBibitem
\bibitem{Liu:2012qua}
C.~Liu and Z.-h. Zhao, ``{$\theta_{13}$ and the Higgs mass from high scale
  supersymmetry},'' \href{http://dx.doi.org/10.1088/0253-6102/59/4/14}{{\em
  Commun. Theor. Phys.} {\bfseries 59} (2013) 467--471},
\href{http://arxiv.org/abs/1205.3849}{{\ttfamily arXiv:1205.3849 [hep-ph]}}.
\mciteBstWouldAddEndPunctfalse
\mciteSetBstMidEndSepPunct{\mcitedefaultmidpunct}
{}{\mcitedefaultseppunct}\relax
\EndOfBibitem
\bibitem{Zheng:2017aaa}
S.~Zheng, ``{Effective Higgs Theories in Supersymmetric Grand Unification},''
  \href{http://dx.doi.org/10.1140/epjc/s10052-017-5174-1}{{\em Eur. Phys. J.}
  {\bfseries C77} no.~9, (2017) 588},
\href{http://arxiv.org/abs/1706.01071}{{\ttfamily arXiv:1706.01071 [hep-ph]}}.
\mciteBstWouldAddEndPunctfalse
\mciteSetBstMidEndSepPunct{\mcitedefaultmidpunct}
{}{\mcitedefaultseppunct}\relax
\EndOfBibitem
\bibitem{Zheng:2019kqu}
S.~Zheng, ``{Minimal Vectorlike Model in Supersymmetric Unification},''
\href{http://arxiv.org/abs/1904.10145}{{\ttfamily arXiv:1904.10145 [hep-ph]}}.
\mciteBstWouldAddEndPunctfalse
\mciteSetBstMidEndSepPunct{\mcitedefaultmidpunct}
{}{\mcitedefaultseppunct}\relax
\EndOfBibitem
\bibitem{Buttazzo:2013uya}
D.~Buttazzo, G.~Degrassi, P.~P. Giardino, G.~F. Giudice, F.~Sala, A.~Salvio,
  and A.~Strumia, ``{Investigating the near-criticality of the Higgs boson},''
  \href{http://dx.doi.org/10.1007/JHEP12(2013)089}{{\em JHEP} {\bfseries 12}
  (2013) 089},
\href{http://arxiv.org/abs/1307.3536}{{\ttfamily arXiv:1307.3536 [hep-ph]}}.
\mciteBstWouldAddEndPunctfalse
\mciteSetBstMidEndSepPunct{\mcitedefaultmidpunct}
{}{\mcitedefaultseppunct}\relax
\EndOfBibitem
\bibitem{Machacek:1983tz}
M.~E. Machacek and M.~T. Vaughn, ``{Two Loop Renormalization Group Equations in
  a General Quantum Field Theory. 1. Wave Function Renormalization},''
\href{http://dx.doi.org/10.1016/0550-3213(83)90610-7}{{\em Nucl. Phys.}
  {\bfseries B222} (1983) 83--103}.
\mciteBstWouldAddEndPunctfalse
\mciteSetBstMidEndSepPunct{\mcitedefaultmidpunct}
{}{\mcitedefaultseppunct}\relax
\EndOfBibitem
\bibitem{PerezLorenzana:1999tf}
A.~Perez-Lorenzana and W.~A. Ponce, ``{GUTs and string GUTs},''
  \href{http://dx.doi.org/10.1209/epl/i2000-00148-y}{{\em Europhys. Lett.}
  {\bfseries 49} (2000) 296--301},
\href{http://arxiv.org/abs/hep-ph/9911540}{{\ttfamily arXiv:hep-ph/9911540
  [hep-ph]}}.
\mciteBstWouldAddEndPunctfalse
\mciteSetBstMidEndSepPunct{\mcitedefaultmidpunct}
{}{\mcitedefaultseppunct}\relax
\EndOfBibitem
\bibitem{Bond:2017wut}
A.~D. Bond, G.~Hiller, K.~Kowalska, and D.~F. Litim, ``{Directions for model
  building from asymptotic safety},''
  \href{http://dx.doi.org/10.1007/JHEP08(2017)004}{{\em JHEP} {\bfseries 08}
  (2017) 004},
\href{http://arxiv.org/abs/1702.01727}{{\ttfamily arXiv:1702.01727 [hep-ph]}}.
\mciteBstWouldAddEndPunctfalse
\mciteSetBstMidEndSepPunct{\mcitedefaultmidpunct}
{}{\mcitedefaultseppunct}\relax
\EndOfBibitem
\bibitem{Nath:2006ut}
P.~Nath and P.~Fileviez~Perez, ``{Proton stability in grand unified theories,
  in strings and in branes},''
  \href{http://dx.doi.org/10.1016/j.physrep.2007.02.010}{{\em Phys. Rept.}
  {\bfseries 441} (2007) 191--317},
\href{http://arxiv.org/abs/hep-ph/0601023}{{\ttfamily arXiv:hep-ph/0601023
  [hep-ph]}}.
\mciteBstWouldAddEndPunctfalse
\mciteSetBstMidEndSepPunct{\mcitedefaultmidpunct}
{}{\mcitedefaultseppunct}\relax
\EndOfBibitem
\bibitem{Miura:2016krn}
{\bfseries Super-Kamiokande} Collaboration, K.~Abe {\em et~al.}, ``{Search for
  proton decay via $p \to e^+\pi^0$ and $p \to \mu^+\pi^0$ in 0.31
  megaton·years exposure of the Super-Kamiokande water Cherenkov detector},''
  \href{http://dx.doi.org/10.1103/PhysRevD.95.012004}{{\em Phys. Rev.}
  {\bfseries D95} no.~1, (2017) 012004},
\href{http://arxiv.org/abs/1610.03597}{{\ttfamily arXiv:1610.03597 [hep-ex]}}.
\mciteBstWouldAddEndPunctfalse
\mciteSetBstMidEndSepPunct{\mcitedefaultmidpunct}
{}{\mcitedefaultseppunct}\relax
\EndOfBibitem
\bibitem{Abe:2018uyc}
{\bfseries Hyper-Kamiokande} Collaboration, K.~Abe {\em et~al.},
  ``{Hyper-Kamiokande Design Report},''
\href{http://arxiv.org/abs/1805.04163}{{\ttfamily arXiv:1805.04163
  [physics.ins-det]}}.
\mciteBstWouldAddEndPunctfalse
\mciteSetBstMidEndSepPunct{\mcitedefaultmidpunct}
{}{\mcitedefaultseppunct}\relax
\EndOfBibitem
\bibitem{Khachatryan:2016mlc}
{\bfseries CMS} Collaboration, V.~Khachatryan {\em et~al.}, ``{Measurement and
  QCD analysis of double-differential inclusive jet cross sections in pp
  collisions at $ \sqrt{s}=8 $ TeV and cross section ratios to 2.76 and 7
  TeV},'' \href{http://dx.doi.org/10.1007/JHEP03(2017)156}{{\em JHEP}
  {\bfseries 03} (2017) 156},
\href{http://arxiv.org/abs/1609.05331}{{\ttfamily arXiv:1609.05331 [hep-ex]}}.
\mciteBstWouldAddEndPunctfalse
\mciteSetBstMidEndSepPunct{\mcitedefaultmidpunct}
{}{\mcitedefaultseppunct}\relax
\EndOfBibitem
\bibitem{Aaboud:2019trc}
{\bfseries ATLAS} Collaboration, M.~Aaboud {\em et~al.}, ``{Search for heavy
  charged long-lived particles in the ATLAS detector in 36.1 fb$^{-1}$ of
  proton-proton collision data at $\sqrt{s} = 13$ TeV},''
  \href{http://dx.doi.org/10.1103/PhysRevD.99.092007}{{\em Phys. Rev.}
  {\bfseries D99} no.~9, (2019) 092007},
\href{http://arxiv.org/abs/1902.01636}{{\ttfamily arXiv:1902.01636 [hep-ex]}}.
\mciteBstWouldAddEndPunctfalse
\mciteSetBstMidEndSepPunct{\mcitedefaultmidpunct}
{}{\mcitedefaultseppunct}\relax
\EndOfBibitem
\bibitem{Farina:2016rws}
M.~Farina, G.~Panico, D.~Pappadopulo, J.~T. Ruderman, R.~Torre, and A.~Wulzer,
  ``{Energy helps accuracy: electroweak precision tests at hadron colliders},''
  \href{http://dx.doi.org/10.1016/j.physletb.2017.06.043}{{\em Phys. Lett.}
  {\bfseries B772} (2017) 210--215},
\href{http://arxiv.org/abs/1609.08157}{{\ttfamily arXiv:1609.08157 [hep-ph]}}.
\mciteBstWouldAddEndPunctfalse
\mciteSetBstMidEndSepPunct{\mcitedefaultmidpunct}
{}{\mcitedefaultseppunct}\relax
\EndOfBibitem
\bibitem{Khachatryan:2016sfv}
{\bfseries CMS} Collaboration, V.~Khachatryan {\em et~al.}, ``{Search for
  long-lived charged particles in proton-proton collisions at $\sqrt s=$ 13
  TeV},'' \href{http://dx.doi.org/10.1103/PhysRevD.94.112004}{{\em Phys. Rev.}
  {\bfseries D94} no.~11, (2016) 112004},
\href{http://arxiv.org/abs/1609.08382}{{\ttfamily arXiv:1609.08382 [hep-ex]}}.
\mciteBstWouldAddEndPunctfalse
\mciteSetBstMidEndSepPunct{\mcitedefaultmidpunct}
{}{\mcitedefaultseppunct}\relax
\EndOfBibitem
\bibitem{Tanabashi:2018oca}
{\bfseries Particle Data Group} Collaboration, M.~Tanabashi {\em et~al.},
  ``{Review of Particle Physics},''
\href{http://dx.doi.org/10.1103/PhysRevD.98.030001}{{\em Phys. Rev.} {\bfseries
  D98} no.~3, (2018) 030001}.
\mciteBstWouldAddEndPunctfalse
\mciteSetBstMidEndSepPunct{\mcitedefaultmidpunct}
{}{\mcitedefaultseppunct}\relax
\EndOfBibitem
\bibitem{Barbieri:2004qk}
R.~Barbieri, A.~Pomarol, R.~Rattazzi, and A.~Strumia, ``{Electroweak symmetry
  breaking after LEP-1 and LEP-2},''
  \href{http://dx.doi.org/10.1016/j.nuclphysb.2004.10.014}{{\em Nucl. Phys.}
  {\bfseries B703} (2004) 127--146},
\href{http://arxiv.org/abs/hep-ph/0405040}{{\ttfamily arXiv:hep-ph/0405040
  [hep-ph]}}.
\mciteBstWouldAddEndPunctfalse
\mciteSetBstMidEndSepPunct{\mcitedefaultmidpunct}
{}{\mcitedefaultseppunct}\relax
\EndOfBibitem
\bibitem{Cacciapaglia:2006pk}
G.~Cacciapaglia, C.~Csaki, G.~Marandella, and A.~Strumia, ``{The Minimal Set of
  Electroweak Precision Parameters},''
  \href{http://dx.doi.org/10.1103/PhysRevD.74.033011}{{\em Phys. Rev.}
  {\bfseries D74} (2006) 033011},
\href{http://arxiv.org/abs/hep-ph/0604111}{{\ttfamily arXiv:hep-ph/0604111
  [hep-ph]}}.
\mciteBstWouldAddEndPunctfalse
\mciteSetBstMidEndSepPunct{\mcitedefaultmidpunct}
{}{\mcitedefaultseppunct}\relax
\EndOfBibitem
\bibitem{Alves:2014cda}
D.~S.~M. Alves, J.~Galloway, J.~T. Ruderman, and J.~R. Walsh, ``{Running
  Electroweak Couplings as a Probe of New Physics},''
  \href{http://dx.doi.org/10.1007/JHEP02(2015)007}{{\em JHEP} {\bfseries 02}
  (2015) 007},
\href{http://arxiv.org/abs/1410.6810}{{\ttfamily arXiv:1410.6810 [hep-ph]}}.
\mciteBstWouldAddEndPunctfalse
\mciteSetBstMidEndSepPunct{\mcitedefaultmidpunct}
{}{\mcitedefaultseppunct}\relax
\EndOfBibitem
\bibitem{Falkowski:2015krw}
A.~Falkowski and K.~Mimouni, ``{Model independent constraints on four-lepton
  operators},'' \href{http://dx.doi.org/10.1007/JHEP02(2016)086}{{\em JHEP}
  {\bfseries 02} (2016) 086},
\href{http://arxiv.org/abs/1511.07434}{{\ttfamily arXiv:1511.07434 [hep-ph]}}.
\mciteBstWouldAddEndPunctfalse
\mciteSetBstMidEndSepPunct{\mcitedefaultmidpunct}
{}{\mcitedefaultseppunct}\relax
\EndOfBibitem
\bibitem{Aad:2016zzw}
{\bfseries ATLAS} Collaboration, G.~Aad {\em et~al.}, ``{Measurement of the
  double-differential high-mass Drell-Yan cross section in pp collisions at $
  \sqrt{s}=8 $ TeV with the ATLAS detector},''
  \href{http://dx.doi.org/10.1007/JHEP08(2016)009}{{\em JHEP} {\bfseries 08}
  (2016) 009},
\href{http://arxiv.org/abs/1606.01736}{{\ttfamily arXiv:1606.01736 [hep-ex]}}.
\mciteBstWouldAddEndPunctfalse
\mciteSetBstMidEndSepPunct{\mcitedefaultmidpunct}
{}{\mcitedefaultseppunct}\relax
\EndOfBibitem
\bibitem{CMS:2014jea}
{\bfseries CMS} Collaboration, V.~Khachatryan {\em et~al.}, ``{Measurements of
  differential and double-differential Drell-Yan cross sections in
  proton-proton collisions at 8 TeV},''
  \href{http://dx.doi.org/10.1140/epjc/s10052-015-3364-2}{{\em Eur. Phys. J.}
  {\bfseries C75} no.~4, (2015) 147},
\href{http://arxiv.org/abs/1412.1115}{{\ttfamily arXiv:1412.1115 [hep-ex]}}.
\mciteBstWouldAddEndPunctfalse
\mciteSetBstMidEndSepPunct{\mcitedefaultmidpunct}
{}{\mcitedefaultseppunct}\relax
\EndOfBibitem
\bibitem{SLANSKY19811}
R.~Slansky, ``Group theory for unified model building,''
  \href{http://dx.doi.org/https://doi.org/10.1016/0370-1573(81)90092-2}{{\em
  Physics Reports} {\bfseries 79} no.~1, (1981) 1 -- 128}.
  \url{http://www.sciencedirect.com/science/article/pii/0370157381900922}\relax
\mciteBstWouldAddEndPunctfalse
\mciteSetBstMidEndSepPunct{\mcitedefaultmidpunct}
{}{\mcitedefaultseppunct}\relax
\EndOfBibitem
\end{mcitethebibliography}
